\documentclass[lettersize,journal]{IEEEtran}

\usepackage{array}
\usepackage[caption=false,font=normalsize,labelfont=sf,textfont=sf]{subfig}
\usepackage{stfloats}
\usepackage{cite}
\usepackage{url}
\usepackage{textcomp}
\usepackage{xcolor}
\usepackage{hyperref}
\usepackage{booktabs}	
\usepackage{diagbox}    
\usepackage{multirow}   
\usepackage{verbatim}	
\usepackage{graphicx}
\usepackage{tabularx}

\usepackage{amsmath,amssymb,amsfonts,graphicx}
\usepackage{epstopdf}

\newtheorem{definition}{Definition}[section]

\usepackage{graphicx,epstopdf,algpseudocode,caption,url}   
\usepackage{graphicx}
\usepackage{caption}
\usepackage{booktabs}

\usepackage[ruled,linesnumbered]{algorithm2e}
\usepackage{amssymb}
\usepackage{hyperref}
\usepackage{bm}
\usepackage[switch]{lineno} 

\begin{document}

\title{Utility-based Privacy Preserving Data Mining}

\author{
	Qingfeng Zhou, Wensheng Gan*,~\IEEEmembership{Member,~IEEE}, Zhenlian Qi, Philip S. Yu,~\IEEEmembership{Life Fellow,~IEEE}  \\
 
        \thanks{This research was supported in part by National Natural Science Foundation of China (No. 62272196), Guangzhou Basic and Applied Basic Research Foundation (No. 2024A04J9971), and the Excellent Graduate Student Cultivation Program of Jinan University. (Corresponding author: Wensheng Gan)}

        \thanks{Qingfeng Zhou and Wensheng Gan are with the College of Cyber Security, Jinan University, Guangzhou 510632, China (E-mail: qingfeng014@gmail.com, wsgan001@gmail.com).}

        \thanks{Zhenlian Qi is with the School of Information Engineering, Guangdong Eco-Engineering Polytechnic, Guangzhou 510520, China. (E-mail: qzlhit@gmail.com)}

        \thanks{Philip S. Yu is with the Department of Computer Science, University of Illinois Chicago, Chicago, IL 60607 USA.  (E-mail: psyu@uic.edu)}
}

\markboth{IEEE INTERNET OF THINGS JOURNAL, 2025}
{}

\maketitle

\begin{abstract}
    With the advent of big data, periodic pattern mining has demonstrated significant value in real-world applications, including smart home systems, healthcare systems, and the medical field. However, advances in network technology have enabled malicious actors to extract sensitive information from publicly available datasets, posing significant threats to data providers and, in severe cases, hindering societal development. To mitigate such risks, privacy-preserving utility mining (PPUM) has been proposed. However, PPUM is unsuitable for addressing privacy concerns in periodic information mining. To address this issue, we innovatively extend the existing PPUM framework and propose two algorithms, Maximum sensitive Utility-MAximum maxPer item (MU-MAP) and Maximum sensitive Utility-MInimum maxPer item (MU-MIP). These algorithms aim to hide sensitive periodic high-utility itemsets while generating sanitized datasets. To enhance the efficiency of the algorithms, we designed two novel data structures: the Sensitive Itemset List (SISL) and the Sensitive Item List (SIL), which store essential information about sensitive itemsets and their constituent items. Moreover, several performance metrics were employed to evaluate the performance of our algorithms compared to the state-of-the-art PPUM algorithms. The experimental results show that our proposed algorithms achieve an Artificial Cost (AC) value of 0 on all datasets when hiding sensitive itemsets. In contrast, the traditional PPUM algorithm yields non-zero AC. This indicates that our algorithms can successfully hide sensitive periodic itemsets without introducing misleading patterns, whereas the PPUM algorithm generates additional itemsets that may interfere with user decision-making. Moreover, the results also reveal that our algorithms maintain Database Utility Similarity (DUS) of over 90\% after the sensitive itemsets are hidden. The code and datasets are publicly available at \href{https://github.com/DSI-Lab1/PPUM}{https://github.com/DSI-Lab1/PPUM.}   
\end{abstract}


\section{Introduction} \label{sec:introduction}

With the development of large-scale databases and the proliferation of the internet, merely collecting and storing vast amounts of data is no longer sufficient to unlock its potential value. Instead, knowledge and insights should be extracted from the rich data to achieve this purpose. Consequently, data mining \cite{rage2024pami,jazayeri2024frequent,gao2023toward,li2024opf} emerged to reveal hidden patterns and trends within data. As a branch of data mining, utility-driven pattern discovery \cite{gan2021survey,wu2021haop}, such as high-utility itemset mining  (HUIM) \cite{tseng2010UP-Growth,liu2012mining}, enables users to extract valuable information from large datasets. HUIM focuses on extracting itemsets from a dataset with utility values greater than or equal to a minimum utility threshold ($\varphi$). Existing HUIM algorithms include  Two-Phase \cite{liu2005two}, UP-Growth \cite{tseng2010UP-Growth}, HUI-Miner \cite{liu2012mining}, FHM \cite{fournier2014fhm}, EFIM \cite{zida2015efim}, DPHIM \cite{2023HUIOnMulticoreProcessors}, R-Miner \cite{sra2024residual}, EMHUN \cite{tung2024efficient}, and FOTH \cite{yan2024efficient}. HUIM has significant practical applications, as it considers both the frequency of itemsets in a dataset and their utility (e.g., profit, weight, and satisfaction), enabling more informed decision-making. In web clickstream analysis \cite{gan2021survey}, HUIM effectively captures user behavior and preferences, improving the user experience and supporting targeted marketing strategies. Additionally, in cross-selling \cite{tseng2012efficient}, HUIM identifies product relationships, helping boost sales and improve customer satisfaction.

The results produced by utility-driven data mining may have substantial value in the short term but may not retain high value over time. For instance, certain products capitalize on customer curiosity, achieving high sales in the short run but experiencing a decline in sales over the long term. To address this, periodic high-utility itemset mining (PHUIM) \cite{zhou2025targeted,iqbal2024mining} was developed to uncover information within databases that retains substantial value over extended periods. PHUIM is designed to identify itemsets in a database that meet the user-defined periodic thresholds and have utility values no less than $\varphi$. The periodicity constraints primarily consist of four parameters: minimum period (\textit{minPer}), maximum period (\textit{maxPer}), minimum average period (\textit{minAvg}), and maximum average period (\textit{maxAvg}). Existing PHUIM algorithms primarily include PHM \cite{fournier2016phm}, PHMN \cite{lai2023PHMN}, and others. According to previous studies, periodic pattern mining has significant real-world applications (smart city management \cite{ma2023understanding}, telehealth systems \cite{ismail2018mining}, and medical fields \cite{huang2014online}).

\begin{figure}[h]
  \centering
  \includegraphics[clip,scale=0.3]{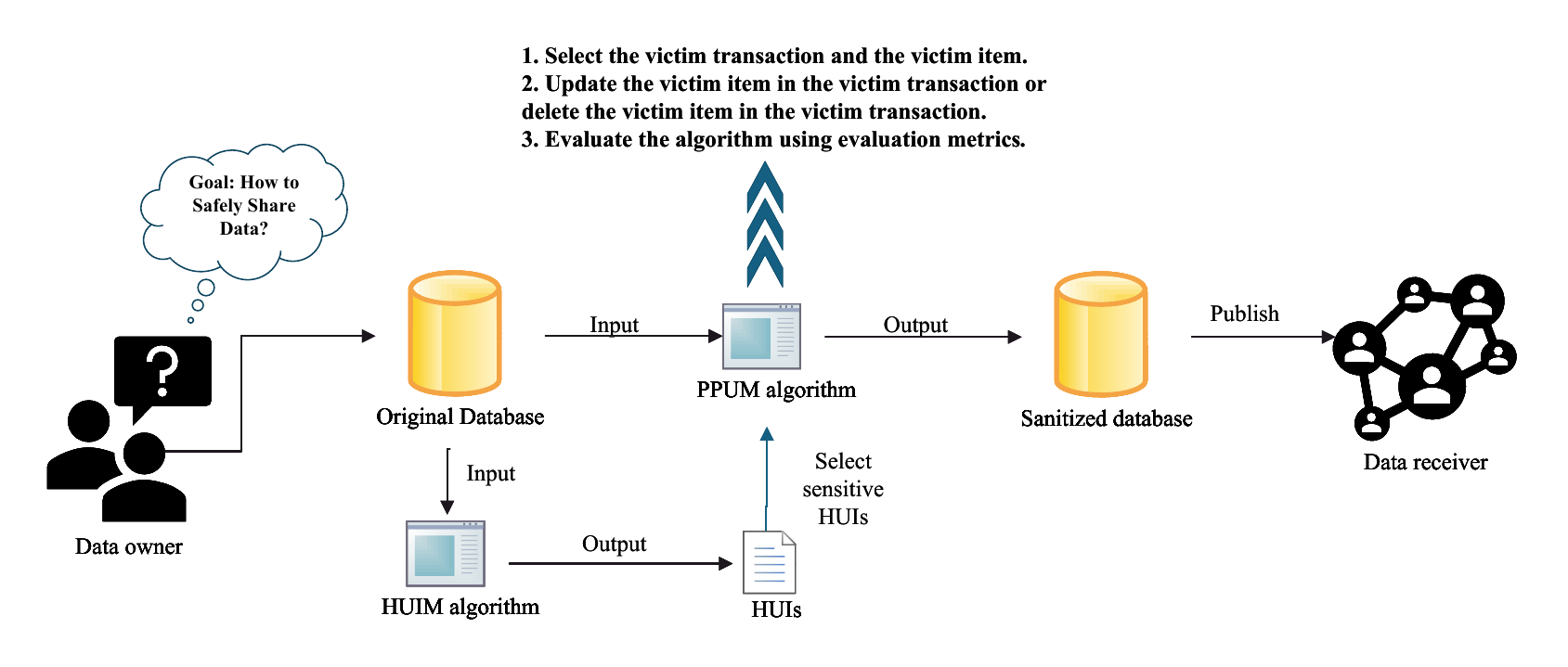} 
  \caption{The workflow of privacy-preserving utility mining.}
  \label{fig:workflowofppum}
\end{figure}

With the widespread use of the internet, collaboration between individuals has intensified, making data and resource sharing a crucial driver of collective progress. However, security risks in online environment have enabled malicious actors to extract sensitive information from datasets, such as customers' names, ID numbers, and home addresses. This will compromise customer privacy and cause adverse effects. Therefore, privacy-preserving utility mining (PPUM) algorithms \cite{gan2018privacy,lin2016fast}  have been developed. These algorithms allow data owners to preprocess datasets before sharing them, preventing them from exposing sensitive information and enabling them to share those processed datasets securely. The workflow of PPUM is illustrated in Fig. \ref{fig:workflowofppum}. Notable classic PPUM algorithms include HHUIF \cite{yeh2010hhuif}, MSICF \cite{yeh2010hhuif}, FPUTT \cite{yun2015fast}, MSU-MAU \cite{lin2016fast}, MSU-MIU \cite{lin2016fast}, MinMax and Weighted \cite{jangra2022efficient}, SMAU \cite{liu2020effective}, SMIU \cite{liu2020effective}, and SMSE \cite{liu2020effective}. HHUIF \cite{yeh2010hhuif} modifies items with the highest utility in sensitive transactions, while MSICF \cite{yeh2010hhuif} targets items with the highest conflict count. FPUTT \cite{yun2015fast} improves efficiency through a compact data structure, reducing database scans. MSU-MAU and MSU-MIU \cite{lin2016fast} select transactions based on maximum sensitive utility, differing in whether they modify the item with maximum or minimum utility. The MinMax and Weighted algorithms \cite{jangra2022efficient} consider side effects on non-sensitive and other sensitive itemsets during sanitization. SMAU, SMIU, and SMSE \cite{liu2020effective} prioritize transactions with fewer non-sensitive itemsets, using different criteria—maximum utility, minimum utility, or minimal side effects—for item modification. Despite sharing the common principle of adjusting sensitive item quantities, these methods differ in their optimization goals and selection strategies.

PPUM is highly important in various practical applications. For instance, in retail and e-commerce, merchants analyze customer purchase records to optimize inventory and marketing strategies. However, existing PPUM methods often fail to consider the periodicity of itemsets, even though many high-utility patterns in real-life scenarios exhibit distinct periodic characteristics. For example, in the healthcare domain, patients' periodic visit patterns need to be concealed to protect privacy; in the finance and insurance sectors, clients' periodic transaction behaviors may involve sensitive financial information. Therefore, integrating privacy-preserving technologies to protect PHUIs is crucial for balancing data value and privacy protection. There exist various privacy-preserving techniques for data protection, such as encryption, anonymization, data perturbation, and data sanitization. Encryption and anonymization safeguard user information by transforming it into an unreadable or unidentifiable format. Data perturbation, on the other hand, protects user privacy by introducing noise into the data, thereby preventing attackers from extracting sensitive information. In this study, we focus on a privacy-preserving technique based on data sanitization, which modifies the original data directly to prevent unauthorized access to sensitive information. The approach adopted in this work belongs to a fundamentally different framework from those of encryption, anonymization, and data perturbation.

As the demand for discovering interesting periodic patterns from databases has increased, the need to share data securely without disclosing sensitive periodic information has become crucial. Traditional PPUM algorithms, such as SMAU, SMIU, and SMSE \cite{liu2020effective}, do not adequately account for the periodic nature of itemsets, which may result in changes in itemset periodicity when items are removed from the database. Direct application of these techniques to sanitize sensitive periodic information may lead to the generation of AC, which can mislead users and compromise the accuracy of decision-making. Consequently, conventional PPUM algorithms are not suitable for preserving the privacy of periodic information. To tackle the above issues, we formulate a novel problem named \textbf{P}rivacy-\textbf{P}reserving \textbf{P}eriodic high-\textbf{U}tility itemset \textbf{M}ining (PPPUM). Leveraging the foundations of the existing PPUM framework and PHUIM techniques, we design two tailored data structures and propose two dedicated algorithms to address this problem efficiently and accurately. The proposed algorithm achieves a balance between protecting sensitive itemsets and maintaining the periodic nature of itemsets, without generating any AC. The main contributions of this paper are summarized as follows.

\begin{itemize}
    \item It introduces the PPPUM problem and proposes two algorithms, Maximum sensitive Utility-MAximum maxPer item (MU-MAP) and Maximum sensitive Utility-MInimum maxPer item (MU-MIP), to address this challenge effectively.

    \item  To enhance algorithm efficiency, two novel data structures, Sensitive Itemset List (SISL) and Sensitive Item List (SIL), are developed to store critical information about sensitive itemsets and their corresponding items.

    \item  We evaluated the performance of MU-MAP and MU-MIP against the state-of-the-art PPUM algorithms (SMAU, SMIU, and SMSE) on six real-world datasets. The experimental results demonstrate that our proposed algorithms can successfully and efficiently hide sensitive periodic itemsets, whereas the PPUM algorithms introduce artificial costs (AC) when hiding sensitive periodic itemsets.
\end{itemize}

The paper is structured as follows: Section \ref{sec: relatedwork} reviews previous studies on HUIM, PHUIM, and PPUM. Section \ref{sec: preliminaries} introduces key concepts related to PPPUM. Section \ref{sec: algorithm} provides a detailed explanation of the MU-MAP and MU-MIP algorithms. Section \ref{sec: experiments} discusses the experimental setup, results, and analysis. Finally, Section \ref{sec: conclusion} summarizes the findings and outlines potential avenues for future research.

\section{Related Work} \label{sec: relatedwork}

This section presents an overview of the related work on HUIM, PHUIM, and PPUM.

\subsection{Utility mining}

High-utility itemset mining (HUIM) \cite{liu2023mining,luna2023efficient,nguyen2023parallel} aims to identify patterns in databases with utility values not less than a predefined threshold ($\varphi$). However, utility in itemsets does not satisfy the downward closure property; an itemset’s utility can be higher, equal to, or lower than its supersets or subsets. To overcome this limitation, various HUIM algorithms have been proposed. The Two-Phase algorithm \cite{liu2005two} aims to reduce the number of candidate itemsets efficiently and identify all high-utility itemsets by introducing the transaction-weighted utilization (TWU) model and the transaction-weighted downward closure property (TWDCP). These innovations limit the search space and ensure coverage of all high-utility itemsets, requiring only a single additional database scan to filter out overestimated itemsets. The UP-Growth and UP-Growth+ algorithms \cite{tseng2012UP-GrowthAndUP-Growth+} are designed to extract high-utility itemsets from transactional databases. These algorithms utilize a data structure called UP-Tree \cite{han2000FP-Growth}, which stores high-utility itemset information and efficiently generates potential high-utility itemsets with only two database scans. The HUI-Miner algorithm \cite{liu2012mining} can discover high-utility itemsets without generating candidates, thus avoiding the computational cost of candidate generation and utility computation. Its novelty lies in the introduction of the utility list data structure, which not only stores utility information but also includes heuristic information to determine if an itemset should be pruned. HUI-Miner employs a new pruning strategy called remaining utility pruning to improve mining efficiency. Once an initial utility list is created, HUI-Miner can mine high-utility itemsets directly from these lists. The FHM algorithm \cite{fournier2014fhm} integrates a novel strategy called EUCP, enabling itemset pruning without performing joins, thereby reducing join operations when mining high-utility itemsets using utility list data structures. The EFIM algorithm \cite{zida2015efim} introduces two novel upper bounds, sub-tree utility and the local utility, for pruning the search space. It also proposes a unique array-based utility counting method, called FUC, for linear time and space computation of these bounds. To further reduce database scanning costs, EFIM implements HDP and HTM, which can also be achieved in linear time and space. The FHN algorithm \cite{lin2016fhn} features a vertical list structure, the PNU-list, designed to efficiently mine high-utility itemsets with negative unit profits. To address the limitations of utility list-based techniques, a new algorithm, R-Miner \cite{sra2024residual}, has been proposed. R-Miner relies on two innovative data structures, the residue map and the master map, to improve the mining efficiency of the high-utility itemset. The FOTH algorithm \cite{yan2024efficient} utilizes a novel structure known as IndexSet to extract HUIs, eliminating the need for join operations through the propagation of IndexSets. To address the challenges of mining HUIs in large-scale datasets, a GPU-based heuristic algorithm called PHA-HUIM \cite{fang2023gpu} has been developed; a parallelized method named DPHIM \cite{2023HUIOnMulticoreProcessors} has been introduced, offering an efficient solution for HUI mining. While HUIM focuses on identifying HUIs that meet $\varphi$, it does not sufficiently consider two critical aspects: the periodicity of itemsets and the protection of user privacy. This work aims to bridge this gap by integrating periodicity constraints into the mining process while ensuring privacy preservation.

\subsection{Periodic pattern mining}

The aforementioned algorithms demonstrate effectiveness in HUIM; however, focusing exclusively on itemset utility during mining can be restrictive. This limitation arises because itemsets with high short-term utility may lack sustained relevance over extended periods. To address this, the FP2M and SFP2M algorithms \cite{zhang2022fuzzy} were developed to identify periodic frequent patterns. Furthermore, Fournier-Viger et al. \cite{fournier2016phm} introduced the concept of PHUIM and created the PHM algorithm, which aims to uncover itemsets that maintain high utility over longer durations by considering both utility and periodicity within transaction databases. The PHM algorithm uses an UL structure \cite{liu2012mining,fournier2014fhm} for mining. For improved efficiency, the algorithm incorporates two pruning strategies—\textit{maxPer} pruning and \textit{maxAvg} pruning—alongside two optimization strategies, ALC pruning and EAPP pruning, based on the ESCS structure. The RSPHUIM algorithm \cite{kenny2023rsphuim} was developed to mine recent short-period HUIs. The TMPHP algorithm \cite{zhou2025targeted} was developed to mine the targeted PHUIs. In a notable advancement, Qi et al. \cite{qi2023mining} proposed the CPR-Miner algorithm, the first algorithm dedicated to discovering closed periodic recent HUIs. Lai et al. \cite{lai2023PHMN} developed the PHMN algorithm to discover potential HUIs containing negative items. PHMN’s innovation lies in its introduction of the M-list structure, a new upper bound, and the DU strategy. These improvements further reduce the search space, leading to the enhanced PHMN+ algorithm with refined DU pruning. While PHUIM extends HUIM by integrating periodic constraints into the mining process, it still does not sufficiently consider an important aspect: the protection of user privacy. Our goal is to address this gap by developing methods that can discover PHUIs without compromising sensitive user information.

\subsection{Privacy-preserving utility mining}

To achieve the goal of hiding sensitive itemsets, Yeh et al. \cite{yeh2010hhuif} proposed a privacy-preserving utility mining model and designed the HHUIF and MSICF algorithms. These algorithms prevent adversaries from extracting sensitive information from the modified database by reducing the utility values of transactions containing sensitive items below a preset threshold while also preventing the reconstruction of the original database. Yun et al. \cite{yun2015fast} introduced a fast PPUM algorithm called FPUTT, which addresses issues in previous PPUM methods by utilizing the FPUTT tree and employing two data structures, the SI table and II table, to enhance perturbation effects. Lin et al. \cite{lin2016fast} developed two algorithms, MSU-MAU and MSU-MIU, to efficiently delete sensitive high-utility itemsets (SHUI) or reduce their utility to achieve privacy protection. A projection mechanism was employed to accelerate the execution of these algorithms. Since the three side effects used to evaluate PPDM were insufficient to assess the performance of PPUM algorithms comprehensively, three new criteria: Itemset Utility Similarity (IUS), Database Utility Similarity (DUS), and Database Structure Similarity (DSS), are introduced to provide a clearer comparison of the efficiency and effectiveness of the developed PPUM algorithms. For datasets containing multiple sensitive patterns, Jangra et al. \cite{jangra2022efficient} proposed two PPUM algorithms: MinMax and Weighted. These algorithms differ in how they select victim items, and each algorithm has three variations based on different methods for selecting sensitive transactions. Gui et al. \cite{gui2024privacy} introduced a new task in privacy-preserving rare pattern mining, designing the LT-MIN and LT-MAX algorithms to hide sensitive rare itemsets and generate a sanitized database while leveraging a projection mechanism to enhance the efficiency of the algorithms. IMSICF \cite{liu2020improved} achieves PPUM by identifying the victim item with the highest conflict count. Liu et al. \cite{liu2020effective} proposed the SMAU, SMIU, and SMSE algorithms to hide all sensitive itemsets while minimizing the side effects of the sanitization process on non-sensitive knowledge. The ODTT algorithm \cite{gui2023odtt} guarantees differential privacy protection for set-valued data. DP-PartFIM \cite{liu2024dp} is an efficient partition-based mining technique leveraging differential privacy, enabling data mining while safeguarding privacy. While PPUM is designed to protect sensitive user information during utility mining, it does not sufficiently consider the temporal or periodic nature of such data. Our objective is to develop privacy-preserving techniques specifically for periodic sensitive information, which requires the integration of both privacy protection and periodicity analysis.

\section{Preliminaries} \label{sec: preliminaries}

Let the transaction dataset $\mathcal{D}$ = \{$T_{1}$, $T_{2}$, $T_{3}$, $\cdots$, $T_{m}$\} represent a series of transactions. Let $I$ = \{ $i_{1}$, $i_{2}$, $i_{3}$, $\cdots$, $i_{n}$ \} represent the set of all items that appear in $\mathcal{D}$. If an itemset $X$ consists of $k$ distinct items, it is called an $k$-itemset. Each transaction has a unique identifier $id$, a set of items \textit{IS}, and their quantities, where 1 $\le$ $id$ $\le$ $m$, and $m$ is the total number of transactions in $\mathcal{D}$, and \textit{IS} $\subseteq$ $I$. The quantity of item $i_{\varsigma}$ in the transaction $T_{id}$ is denoted as \textit{iu}$(i_{\varsigma}, T_{id})$. Let \textit{UTable} represent the utility table, where the external utility of each item is recorded. The external utility of the item $i_{\varsigma}$ is represented by \textit{eu}$(i_{\varsigma})$. For example, the transaction dataset in Table \ref{Database} consists of 10 transactions. The transaction $T_{9}$ contains three distinct items: $c$, $d$, and $e$, with quantities \textit{iu}$(c, T_{9})$ = 5, \textit{iu}$(d, T_{9})$ = 7, and \textit{iu}$(e, T_{9})$ = 8. Table \ref{tab:external_utility} shows the external utility values of all items in $\mathcal{D}$, where \textit{eu}$(a)$ = 8, \textit{eu}$(b)$ = 7, \textit{eu}$(c)$ = 3, \textit{eu}$(d)$ = 7, \textit{eu}$(e)$ = 4, \textit{eu}$(f)$ = 3, \textit{eu}$(g)$ = 2, \textit{eu}$(h)$ = 4, and \textit{eu}$(i)$ = 10.

\begin{table}[!h]
    \footnotesize
    \centering
    \caption{An example database}
    \begin{tabular}{|c|c|c|}
    \hline
    \textbf{Tid} & \textbf{Transaction\{item: quantity\}} & \textbf{TU} \\
    \hline
    $T_{1}$ & \{$a$:6\} \{$b$:5\} \{$e$:5\} \{$g$:6\} \{$h$:9\} \{$i$:2\} & \$171 \\
    \hline
    $T_{2}$ & \{$c$:2\} \{$d$:2\} \{$e$:3\} \{$f$:9\} \{$g$:3\} &  \$65   \\
    \hline
    $T_{3}$ &  \{$b$:8\} \{$c$:5\} \{$f$:3\}  & \$80    \\
    \hline
    $T_{4}$ & \{$a$:9\} \{$b$:6\} \{$d$:5\} \{$e$:1\} \{$f$:2\} \{$h$:7\} & \$187 \\
    \hline
    $T_{5}$ & \{$a$:8\} \{$d$:1\} \{$g$:4\} \{$i$:9\} &  \$169   \\
    \hline
    $T_{6}$ &  \{$a$:3\} \{$b$:1\} \{$c$:5\} \{$d$:1\} \{$e$:6\} \{$f$:2\} \{$g$:6\} \{$i$:2\} & \$115    \\
    \hline
    $T_{7}$ & \{$a$:5\} \{$g$:5\} \{$h$:5\} & \$70 \\
    \hline
    $T_{8}$ & \{$a$:7\} \{$b$:9\} \{$c$:6\} \{$d$:1\} \{$f$:6\} \{$g$:5\} \{$i$:2\} & \$192  \\
    \hline
    $T_{9}$ & \{$c$:5\} \{$d$:7\} \{$e$:8\} & \$96  \\
    \hline
    $T_{10}$ & \{$a$:1\} \{$b$:2\} \{$c$:8\} \{$h$:1\} \{$i$:9\} & \$140  \\
    \hline
    \end{tabular}
    \label{Database}
\end{table}

\begin{table}[h]
    \footnotesize
    \centering
    \caption{Utility table}
    \begin{tabular}{|c|c|c|c|c|c|c|c|c|c|}
        \hline
        \textbf{Item} & $a$ & $b$ & $c$ & $d$ & $e$ & $f$ & $g$ & $h$ & $i$ \\
        \hline
        \textbf{Utility (\$)} & 8 & 7 & 3 & 7 & 4 & 3 & 2 & 4 & 10 \\
        \hline
        \end{tabular}
    \label{tab:external_utility}
\end{table}

\begin{definition}[utility]
    \rm The utility of an item $i_{\varsigma}$ in a transaction $T_{id}$ is defined as $U(i_{\varsigma}, T_{id})$ = \textit{iu}$(i_{\varsigma}, T_{id})$ $\times$ \textit{eu}$(i_{\varsigma})$, where \textit{iu}$(i_{\varsigma}, T_{id})$ represents the internal utility of the item $i_{\varsigma}$ in the transaction $T_{id}$, and \textit{eu}$(i_{\varsigma})$ represents the external utility of item $i_{\varsigma}$. The utility of an itemset $X$ in a transaction $T_{id}$ is defined as: $U(X, T_{id})$ = $\sum_{g \in X \land X\subseteq T_{id} }$ $U(g, T_{id})$. The utility of an itemset $X$ in $\mathcal{D}$ is given by: $U(X)$ = $\sum_{X \subseteq T_{id} \subseteq \mathcal{D}}$ $U(X,T_{id})$. This definition captures the utility calculation for an item and an itemset both within a transaction and across the entire database.
\end{definition}

For instance, $U(a, T_{1})$ = \textit{iu}$(a, T_{1})$ $\times$ \textit{eu}$(a)$ = 6 $\times$ 8 = 48, $U(i, T_{1})$ = \textit{iu}$(i, T_{1})$ $\times$ \textit{eu}$(i)$ = 2 $\times$ 10 = 20, $U(\{a,i\}, T_{1})$ = $\sum_{g \in \{a,i\} \land \{a,i\}\subseteq T_{1} }$ $U(g, T_{1})$ = $U(a, T_{1})$ + $U(i, T_{1})$ = 48 + 20 = 68. Thus, $U(\{a,i\})$ = $\sum_{\{a,i\} \subseteq T_{id} \subseteq \mathcal{D}}$ $U(\{a,i\},T_{id})$ = $U(\{a,i\},T_{1})$+ $U(\{a,i\},T_{5})$ + $U(\{a,i\},T_{6})$ + $U(\{a,i\},T_{8})$ + $U(\{a,i\},T_{10})$ = 68 + 154 + 44 + 76 + 98 = 440.

\begin{definition}[period]
    \rm In $\mathcal{D}$, the set of transaction identifiers containing itemset $X$ is denoted as \textit{ID}$(X)$ = $\{ id | X \subseteq T_{id}\}$ = \{$id_{1}$, $id_{2}$, $\cdots$, $id_{l}$\}, where $l$ $\leqslant$ $|\mathcal{D}|$. Here, $l$ represents the number of transactions containing $X$ and $|\mathcal{D}|$ is the total number of transactions in $\mathcal{D}$. Based on \textit{ID}$(X)$, we can define the periodicity of $X$ as $P(X)$ = \{ $tid_{j+1}$ - $tid_{j}$ $|$ $j$ = 0, 1, 2, $\cdots$, $l$ \}, where $tid_{0}$ = 0 and $tid_{l+1}$ = $|\mathcal{D}|$. From $P(X)$, the maximum, the minimum, and the average periodicity of $X$ can be obtained: \textit{maxPer}$(X)$ = \textit{max(P(X))}, \textit{minPer}$(X)$ = \textit{min(P(X))}, and \textit{avgPer}$(X)$ = $\frac{ \sum_{j \in P(X)} j }{|P(X)|} $ = $\frac{|\mathcal{D}|}{ | sp(X) | + 1} $. Here, $|$\textit{sp(X)}$|$ represents the support of $X$. When calculating \textit{minPer}$(X)$, we exclude the first and last elements of $P(X)$. The detailed reasoning for this exclusion is thoroughly explained in the literature \cite{fournier2016phm}.
\end{definition}

\begin{table}[!h]
    \footnotesize
    \centering
    \renewcommand{\arraystretch}{1.5}
    \setlength{\tabcolsep}{3pt}
    \caption{The period of all single items.}
    \begin{tabular}{|c|c|c|c|c|c|c|}
    \hline
    \textbf{item} & \textbf{ID} & \textbf{Periodicity} & \textbf{\textit{minPer}} & \textbf{\textit{maxPer}} & \textbf{\textit{avgPer}} & \textbf{\textit{sp}} \\
    \hline
    $a$ & 1,4,5,6,7,8,10 & 1,3,1,1,1,1,2,0 & 1 &3& 1.25 & 7 \\
    \hline
    $b$ & 1,3,4,6,8,10 & 1,2,1,2,2,2,0 & 1 &2& 1.43 & 6 \\
    \hline
    $c$ & 2,3,6,8,9,10 & 2,1,3,2,1,1,0 & 1 &3& 1.43 & 6 \\
    \hline
    $d$ & 2,4,5,6,8,9 & 2,2,1,1,2,1,1 & 1 &2& 1.43 & 6 \\
    \hline
    $e$ & 1,2,4,6,9 & 1,1,2,2,3,1 & 1 &3& 1.67 & 5 \\
    \hline
    $f$ & 2,3,4,6,8 & 2,1,1,2,2,2 & 1 &2& 1.67 & 5 \\
    \hline
    $g$ & 1,2,5,6,7,8 & 1,1,3,1,1,1,2 & 1 &3& 1.43 & 6 \\
    \hline
    $h$ & 1,4,7,10 & 1,3,3,3,0 & 3 &3& 2.00 & 4 \\
    \hline
    $i$ & 1,5,6,8,10 & 1,4,1,2,2,0 & 1 &4& 1.67 & 5 \\
    \hline
    \end{tabular}
    \label{tab:ItemPeriodInfo}
\end{table}

For example, in $\mathcal{D}$, the transaction identifiers containing item $e$ are \textit{ID}$(e)$ = \{1, 2, 4, 6, 9\}. Accordingly, the period of item $e$, denoted as $P(e)$, is \{1, 1, 2, 2, 3, 1\}. The maximum period of $e$ is \textit{maxPer}$(e)$ = \textit{max(P(e))} = 3, the minimum period is \textit{minPer(e)} = \textit{min(P(e))} = 1, and the average period is \textit{avgPer(e)} = $\frac{|\mathcal{D}|}{|\textit{sp}(e)| + 1}$ = $\frac{10}{5 + 1}$ = 1.67. The periodicity information of all items in $\mathcal{D}$ is shown in Table \ref{tab:ItemPeriodInfo}.

\begin{definition}[periodic high-utility itemset]
    \rm Let \textit{minPer} represent the minimum period threshold, \textit{maxPer} the maximum period threshold, \textit{minAvg} the minimum average period threshold, and \textit{maxAvg} the maximum average period threshold. If the period of an itemset $X$ is not less than \textit{minPer} and not greater than \textit{maxPer}, the average period of $X$ is between \textit{minAvg} and \textit{maxAvg}, inclusive, and the utility value of $X$ meets or exceeds $\varphi$, then $X$ qualifies as a periodic high-utility itemset. Define \textit{PIs} = \{ $p_{1}$, $p_{2}$, $p_{3}$, $\cdots$, $p_{k}$ \} to represent the set of all periodic high-utility itemsets in $\mathcal{D}$, where each $p_{m}$ $\in$ \textit{PIs} and 1 $\le$ $m$ $\le$ $k$.
\end{definition}

For example, we set \textit{minPer} = 1, \textit{maxPer} = 6, \textit{minAvg} = 1, \textit{maxAvg} = 2, and $\varphi$ = 260. Using data from Table \ref{Database}, we can extract all PHUIs, as shown in Table \ref{tab:PHUIs}.

\begin{table}[!h]
    \footnotesize
    \centering
    \caption{The set of PHUIs for the running example.}
    \begin{tabular}{|l|c|c|c|c|c|}
    \hline
    \textbf{itemset} & \textbf{utility} & \textbf{\textit{sp}} & \textbf{\textit{minPer}} & \textbf{\textit{maxPer}} & \textbf{\textit{avgPer}} \\
    \hline
    \{$a$ $g$ $i$\} & 384 & 4 & 1 & 4 & 2.00 \\ \hline
    \{$a$ $g$\} & 284 & 5 & 1 & 4 & 1.67 \\ \hline
    \{$b$ $i$\} & 269 & 4 & 2 & 5 & 2.00 \\ \hline
    \{$a$ $b$ $i$\} & 405 & 4 & 2 & 5 & 2.00 \\ \hline
    \{$a$ $i$\} & 440 & 5 & 1 & 4 & 1.67 \\ \hline
    \{$a$ $d$\} & 272 & 4 & 1 & 4 & 2.00 \\ \hline
    \{$a$ $b$\} & 369 & 5 & 2 & 3 & 1.67 \\ \hline
    \{$a$\} & 312 & 7 & 1 & 3 & 1.25 \\ \hline
    \end{tabular}
    \label{tab:PHUIs}
\end{table}

\begin{definition}[sensitive itemsets]
    \rm When sharing or publishing data, it is necessary to conceal sensitive information during the data-cleaning process to prevent the disclosure of details such as user ID numbers, names, and home addresses. These sensitive pieces of information are called sensitive itemsets (SIS). Let \textit{SPIs} = \{$sp_{1}$, $sp_{2}$, $sp_{3}$, $\cdots$, $sp_{l}$\} denote the sensitive periodic high-utility itemsets in $\mathcal{D}$, where $sp_{k}$ $\in$ \textit{SPIs}, 1 $\le$ $k$ $\le$ $l$, and \textit{SPIs} $\subseteq$ \textit{PIs}.
\end{definition}

We select the sensitive periodic high-utility itemsets as \{$a$, $b$, $i$\}, \{$a$, $i$\}, and \{$b$, $i$\}. We aim to design an algorithm to hide these itemsets, ensuring that these sensitive pieces of information cannot be mined from the cleaned dataset. Concerning the evaluation criteria in PPDM, we formally define the following metrics in PPPUM. Let \textit{PI} denote the PHUIs in the original database $\mathcal{D}$; \textit{SPI} represents the sensitive PHUIs selected according to user preferences; $\sim$\textit{SPI} indicates the non-sensitive PHUIs; and \textit{PI'} denotes the PHUIs in the sanitized database $\mathcal{D}'$.

\begin{definition}
    \rm  The Hiding Failure (HF), denoted as $\alpha$, is defined by the intersection of \textit{SPI} and \textit{PI'}: $\alpha$ = \textit{SPI} $\cap$ \textit{PI'}. The Missing Cost (MC), denoted as $\beta$, is defined by: $\beta$ = $\sim$\textit{SPI} $-$ \textit{PI'} = (\textit{PI} $-$ \textit{SPI}) $-$ \textit{PI'}. The Artificial Cost (AC), denoted as $\gamma$, is defined by: $\gamma$ = \textit{PI'} $-$ \textit{PI}. HF represents the sensitive itemsets that remain detectable in the sanitized database. MC denotes the non-sensitive PHUIs that were discoverable in the original database but cannot be found in the sanitized version. AC indicates PHUIs appearing in the sanitized database but not in the original database.
\end{definition}

\textbf{Problem statement:} Given a database $\mathcal{D}$ and certain PHUIs constraints (\textit{minPer}, \textit{maxPer}, \textit{minAvg}, \textit{maxAvg}, and $\varphi$), as well as the periodic high-utility itemsets (PHUIs) mined from $\mathcal{D}$ and the selected sensitive periodic high-utility itemsets (SPIs), the task of PPPUM is to use an algorithm to obtain a sanitized database $\mathcal{D}'$. This sanitized $\mathcal{D}'$ should prevent the extraction of these sensitive periodic high-utility itemsets from the dataset.

\section{Proposed Algorithm} \label{sec: algorithm}

In this section, we first propose two novel data structures: the sensitive itemset list and the sensitive item list, to store information about sensitive itemsets. Based on these two data structures, we introduce two new algorithms named Maximum sensitive Utility-MAximum maxPer item (MU-MAP) and Maximum sensitive Utility-MInimum maxPer item (MU-MIP) to hide SPHUIs. Privacy-preserving periodic utility mining consists of three phases: (1) extracting all periodic high-utility itemsets, (2) selecting sensitive periodic high-utility itemsets based on user requirements, and (3) generating a sanitized dataset to prevent the mining of sensitive itemsets. The PHM algorithm \cite{fournier2016phm} is used to mine periodic high-utility itemsets. Since the PHM algorithm in the first phase is well-established and the selection of sensitive itemsets in the second phase involves preprocessing the mining results from the first phase, we will primarily focus on discussing the third phase here: how to hide sensitive itemsets.

\subsection{Data structure}

\begin{definition}[sensitive itemset list]
    \rm For a sensitive itemset $S$ in $\mathcal{D}$, the sensitive itemset list (SISL) for $S$ is composed of five parts: \textit{SIS}, $SU$, \textit{[tid:utility]}, $LP$, and \textit{sup}. Here, \textit{SIS} represents the sensitive itemset, $SU$ denotes the utility value of the sensitive itemset, \textit{[tid:utility]} specifies the transactions containing the sensitive itemset and its utility in each transaction, $LP$ indicates the maximum period of the sensitive itemset, and \textit{sup} represents the support of the sensitive itemset.
\end{definition}

\begin{figure*}[h]
  \centering
  \includegraphics[clip,scale=0.5]{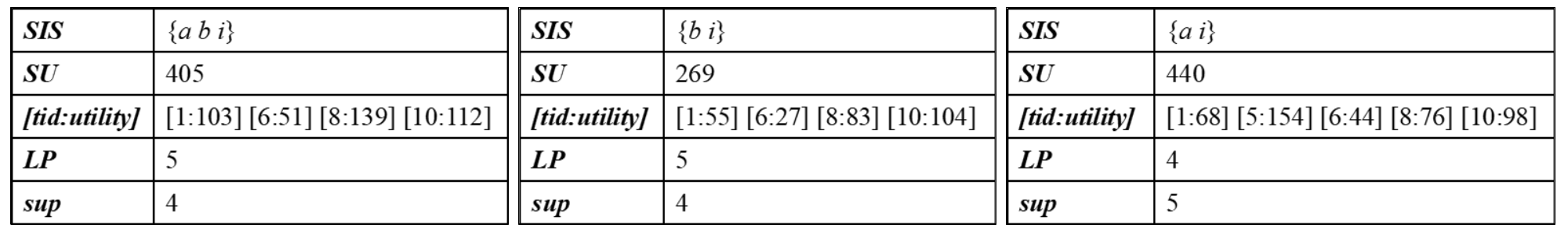} 
  \caption{The SISL for the selected sensitive itemsets.}
  \label{fig:SISL}
\end{figure*}

\begin{definition}[sensitive item list]
    \rm For each sensitive itemset $S$ in $\mathcal{D}$, the sensitive item list (SIL) for $S$ consists of two parts: the sensitive itemset $S$ and a list. This list is a five-tuple, $[$ $tid$, $si$, $iu$, $cnt$, $mp$ $]$, where $tid$ denotes the transaction ID containing $S$, $si$ represents the items within $S$, $iu$ indicates the utility of each item within $S$, $cnt$ specifies the count of items in $S$, and $mp$ stands for the maximum period of the items within $S$.
\end{definition}

\begin{figure}[h]
  \centering
  \includegraphics[clip,scale=0.36]{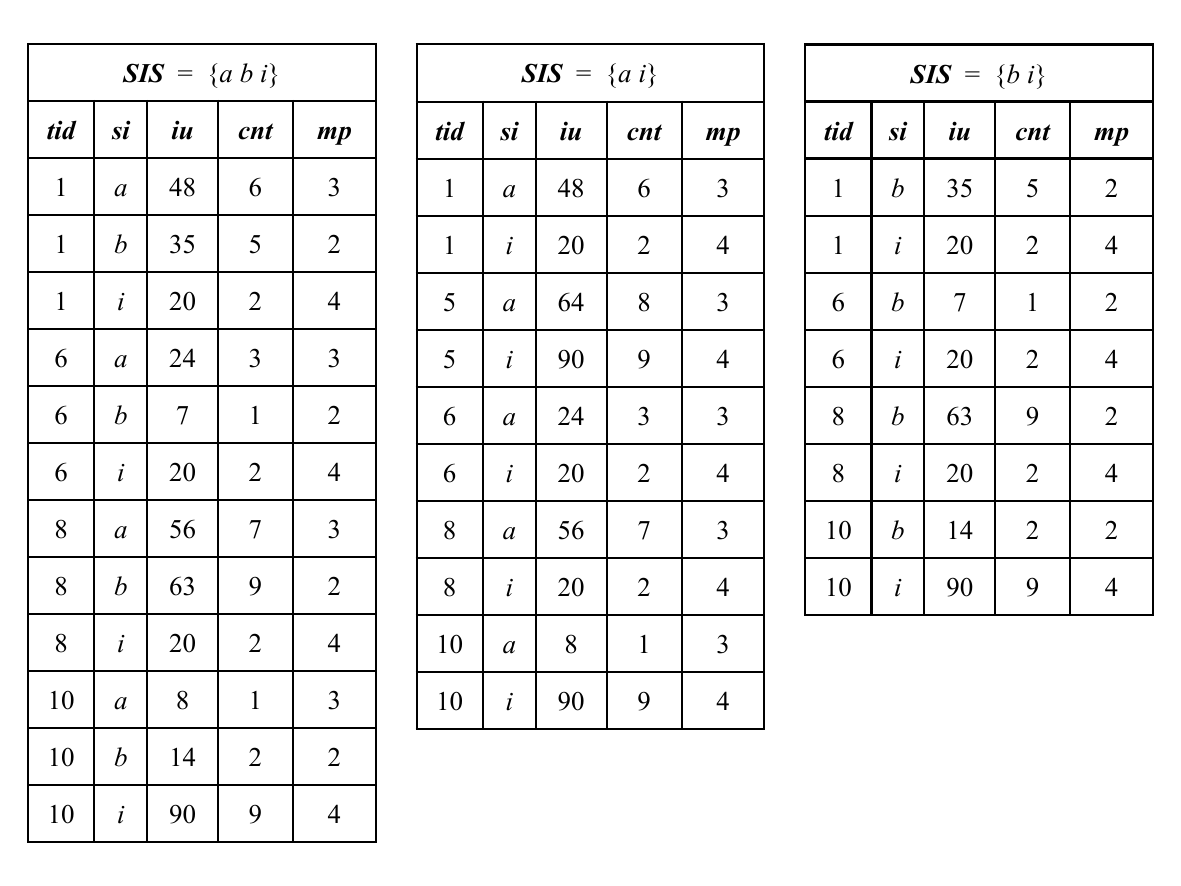} 
  \caption{The SIL for the running example.}
  \label{fig:SIL}
\end{figure}

For example, the SISL and SIL of the selected sensitive itemsets in this study are shown in Fig. \ref{fig:SISL} and Fig. \ref{fig:SIL}, respectively.

\begin{algorithm}[ht]
    \small
    \caption{The MU-MAP algorithm}
    \label{algoMU-MAP}
    \LinesNumbered
    \KwIn{$\mathcal{D}$: the original database; $\varphi$: the minimum utility threshold; \textit{maxPer}: the maximum period threshold; \textit{maxAvg}: the maximum average period threshold; \textit{SPIs}: the sensitive PHUIs in $\mathcal{D}$.}
    \KwOut{$\mathcal{D}'$: the sanitized database.}

    \For{$\forall$ $sp_k$ $\in$  \textit{SPIs} }{
         construct the SISL and SIL structures for $sp_k$\;
    }
    sort each itemset in the \textit{SPIs} in descending order based on the number of items they contain\;
    \textit{minSup} = $|\mathcal{D}|$ / \textit{maxAvg} - 1\;
    \For{$\forall$ $sp_k$ $\in$  \textit{SPIs} }{
       get the SISL and SIL of $sp_k$\;
       $du$($sp_k$) = SISL($sp_k$).$SU$ $-$ $\varphi$\;
       $ds$($sp_k$) = SISL($sp_k$).$sup$\;
       $dp$($sp_k$) = SISL($sp_k$).$LP$\;
       \While{$du$($sp_k$) $\ge$ 0 $\&\&$ $dp$($sp_k$) $\le$ \textit{maxPer} $\&\&$ $ds$($sp_k$) $\ge$ \textit{minSup}}{
          
          $T^{select}$ $\gets$ $\max$ \{ $u$($sp_k$, $T_d$), $T_d$ $\in$ SISL($sp_k$).\textit{[tid:utility]} \}\;

          $i^{select}$ $\gets$  $\max$ \{ $mp$($i_d$, $T^{select}$), 1 $\le$ $d$ $\le$ $|sp_k|$, $i_d$ $\in$ SIL($sp_k$, $T^{select}$) \}\;
   
		 \eIf{SIL($sp_k$, $i^{select}$, $T^{select}$).$iu$ $\le$ $du$($sp_k$)}{
			delete $i^{select}$ from $T^{select}$\;
                SISL($sp_k$).$su$ $-$$=$ $u$($i^{select}$, $T^{select}$)\;
                SISL($sp_k$).$sup$ $-$$-$\;
                SISL($sp_k$).$LP$ $=$ \textit{getNewLP}($sp_k$)\;
                update \textit{SPIs}, SISL and SIL\;
		 }{
			$dq$ = $\lceil$ $du$($sp_k$) $/$ $eu$($i^{select}$) $\rceil$\;
                SIL($i^{select}$, $T^{select}$).$cnt$ $-$$=$ $dq$\;
                SIL($i^{select}$, $T^{select}$).$iu$ $-$$=$ $dq$ $\times$ $eu$($i^{select}$)\;
                delete $sp_k$ from \textit{SPIs}\;
                update SISL and SIL\;
		 }
        }
    }
\end{algorithm}

\begin{figure*}[h]
  \centering
  \includegraphics[clip,scale=0.5]{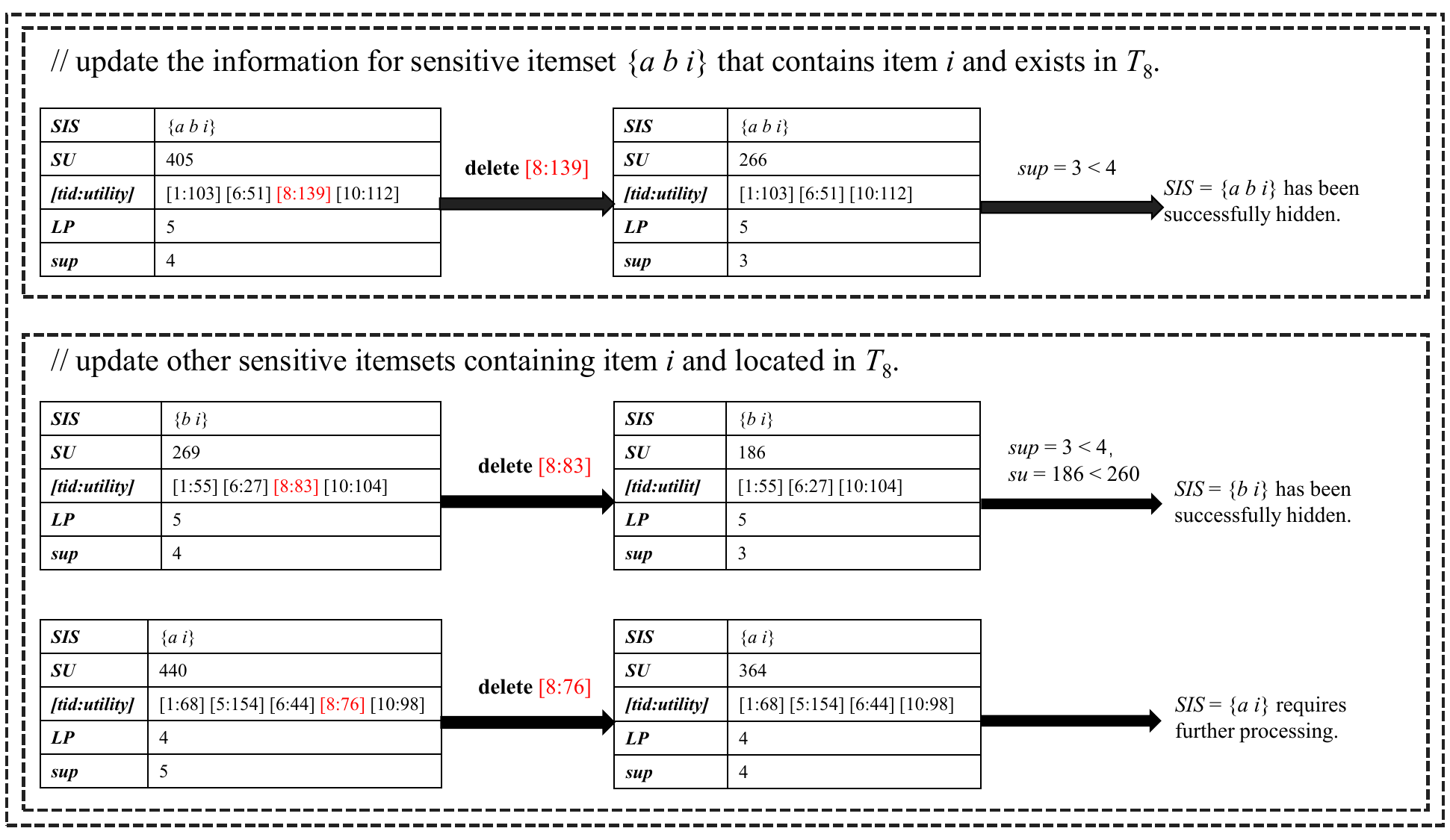} 
  \caption{Remove the item $i$ from the $T_8$ (update SISL).}
  \label{fig:SISLRemove_i_8}
\end{figure*}

\begin{figure*}[h]
  \centering
  \includegraphics[clip,scale=0.5]{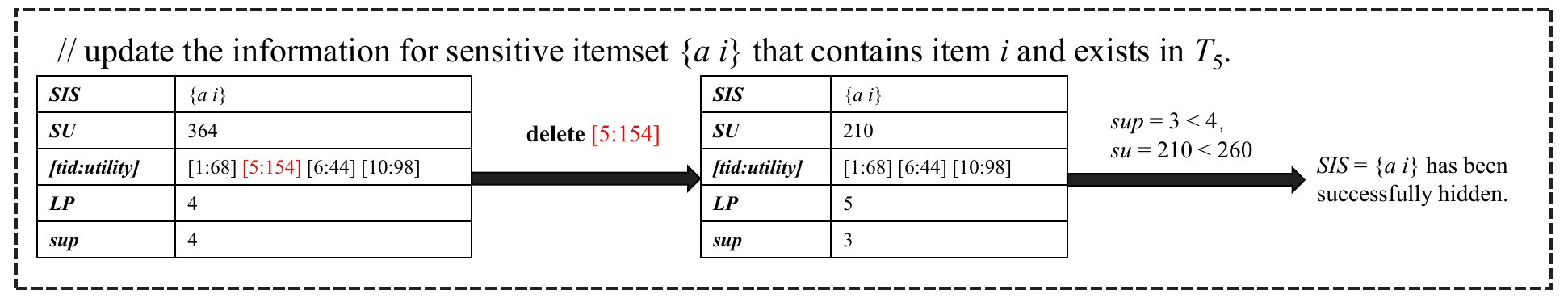} 
  \caption{Remove the item $i$ from the $T_5$ (update SISL).}
  \label{fig:SISLRemove_i_5}
\end{figure*}

\subsection{Proposed MU-MAP algorithm}

The MU-MAP algorithm effectively sanitizes the original database, successfully and efficiently concealing sensitive periodic high-utility itemsets. The algorithm is divided into two main parts: the sensitive periodic high-utility itemset preprocessing phase and the hiding phase. During the preprocessing phase, SISL and SIL are utilized to accelerate the algorithm’s execution time and enhance its efficiency. The detailed process of MU-MAP is presented in Algorithm \ref{algoMU-MAP}.

The first phase is preprocessing SPIs (lines 1-4). Initially, for each sensitive itemset in SPIs, its SISL and SIL are constructed (line 2). Then, the sensitive itemsets are sorted in descending order based on the number of items they contain (line 4). In the algorithm, longer sensitive itemsets are prioritized. The minimum support threshold is calculated using \textit{maxAvg} (line 5). The second phase is the hiding operation on SPIs (lines 6-28). The algorithm iteratively processes each element $sp_k$ in SPIs until all sensitive itemsets are hidden. Each sensitive itemset's SISL and SIL are retrieved (line 7). Using the SISL of the itemset, the threshold values $du$($sp_k$), $ds$($sp_k$), and $dp$($sp_k$) are computed (lines 8-10). Here, $du$($sp_k$) represents the difference between $sp_k$'s utility and $\varphi$; $ds$($sp_k$) denotes $sp_k$'s support, and $dp$($sp_k$) represents $sp_k$'s maximum period. Based on $sp_k$'s SISL, the transaction with the highest utility for $sp_k$ is selected as the victim transaction for modification (line 12). Then, from $sp_k$'s SIL, the item with the highest \textit{maxPer} is selected as the victim item (line 13). The algorithm then decides to delete the item or adjust its quantity in the victim transaction based on its utility (lines 14-27). If the utility of the victim item in the victim transaction does not exceed $du$($sp_k$), the victim item will be removed from the transaction, and SPIs, SISL, and SIL are updated accordingly (lines 14-19). Otherwise, the quantity of the victim item in the victim transaction is reduced (line 22). After this, $sp_k$ is removed from SPIs as it is now hidden, followed by updating SISL and SIL (line 25). Once all SPIs are concealed, the algorithm terminates.

\begin{figure}[h]
  \centering
  \includegraphics[clip,scale=1.1]{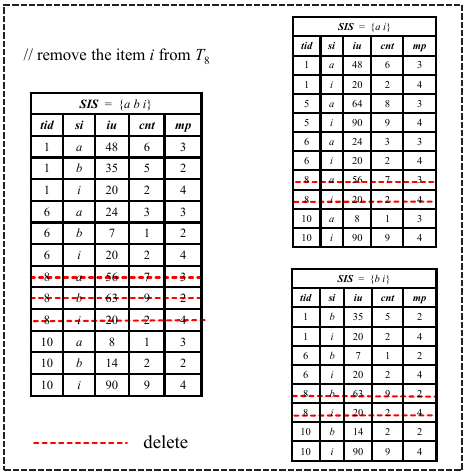} 
  \caption{Remove the item $i$ from $T_8$ (update SIL).}
  \label{fig:SILRemove_i_8}
\end{figure}

\begin{figure}[h]
  \centering
  \includegraphics[clip,scale=0.58]{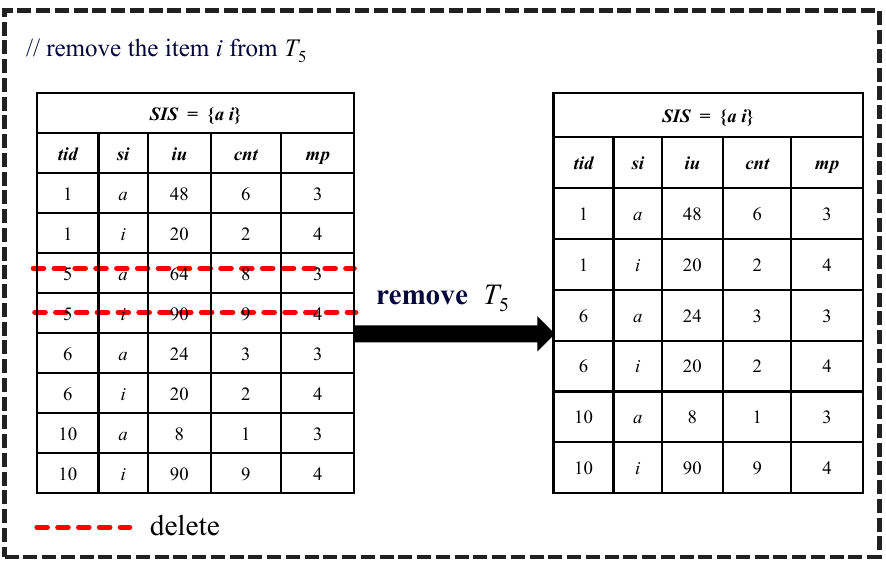} 
  \caption{Remove the item $i$ from $T_5$ (update SIL).}
  \label{fig:SILRemove_i_5}
\end{figure}

Using the selected SPIs and threshold values mentioned earlier, the SPIs are initially sorted based on the number of items they contain, resulting in \{$a$, $b$, $i$\}, \{$a$, $i$\}, and \{$b$, $i$\}. SISL and SIL are then constructed for these itemsets, as shown in Fig. \ref{fig:SISL} and Fig. \ref{fig:SIL}. It starts with \{$a$, $b$, $i$\} since it contains the most items. According to Fig. \ref{fig:SISL}, \{$a$, $b$, $i$\} has the highest utility in $T_8$, with a value of 139. As shown in Fig. \ref{fig:SIL}, the item $i$ has the highest $mp$ in the selected transaction, so the item $i$ is chosen as the victim item. Since $i$ in $T_8$ has a utility of 20, which is less than $du$(\{$a$, $b$, $i$\}) (= 405 $-$ 260 = 145), the item $i$ is deleted from $T_8$. Consequently, the SISL and SIL for all itemsets containing $i$ in $T_8$ are updated, as illustrated in Fig. \ref{fig:SISLRemove_i_8} and Fig. \ref{fig:SILRemove_i_8}. Since the support of both \{$a$, $b$, $i$\} and \{$b$, $i$\} no longer meets $ds$($sp_k$), these itemsets are now fully hidden. The algorithm then proceeds to process \{$a$, $i$\} in the same manner, beginning with $T_5$ and selecting the item $i$ as the victim. The main steps are shown in Fig. \ref{fig:SISLRemove_i_5} and Fig. \ref{fig:SILRemove_i_5}. When all itemsets are hidden, the algorithm terminates.

\subsection{Proposed MU-MIP algorithm}

The MU-MIP algorithm also comprises two main phases: the preprocessing phase for sensitive periodic high-utility itemsets and the hiding phase. The primary distinction between MU-MIP and MU-MAP lies in the selection method for the victim item during the hiding phase. In MU-MIP, items with the smallest \textit{maxPer} are prioritized for handling. The detailed steps of the algorithm are presented in Algorithm \ref{algoMU-MIP}.

The first phase involves preprocessing SPIs (lines 1-4). Initially, for each sensitive itemset in SPIs, its SISL and SIL are constructed (line 2). The sensitive itemsets are then sorted in descending order based on the number of items they contain (line 4), prioritizing longer sensitive itemsets in the algorithm. A minimum support threshold is calculated using \textit{maxAvg} (line 5). The second phase involves the hiding operation on SPIs (lines 6-28). The algorithm iteratively processes each element $sp_k$ in SPIs until all sensitive itemsets are hidden. For each sensitive itemset, its SISL and SIL are retrieved (line 7). Using the itemset's SISL, threshold values $du$($sp_k$), $ds$($sp_k$), and $dp$($sp_k$) are computed (lines 8-10), where $du$($sp_k$) is the difference between $sp_k$'s utility and $\varphi$, $ds$($sp_k$) represents $sp_k$'s support, and $dp$($sp_k$) indicates $sp_k$'s maximum period. Based on $sp_k$'s SISL, the transaction with the highest utility for $sp_k$ is selected as the victim transaction for modification (line 12). Then, the item with the smallest \textit{maxPer} is chosen as the victim item from $sp_k$'s SIL (line 13). The algorithm then determines whether to delete the item or adjust its quantity in the victim transaction, depending on the utility (lines 14-27). If the utility of the victim item in the transaction is less than or equal to $du$($sp_k$), the item is removed from the transaction, and SPIs, SISL, and SIL are updated (lines 14-19). Otherwise, the quantity of the victim item is reduced (line 22). Subsequently, $sp_k$ is removed from SPIs as it is now hidden, and SISL and SIL are updated (line 25). The algorithm concludes once all SPIs are concealed.

\begin{algorithm}[ht]
    \small
    \caption{The MU-MIP algorithm}
    \label{algoMU-MIP}
    \LinesNumbered
    \KwIn{$\mathcal{D}$: the original database; $\varphi$: the minimum utility threshold; \textit{maxPer}: the maximum period threshold; \textit{maxAvg}: the maximum average period threshold; \textit{SPIs}: the sensitive PHUIs in $\mathcal{D}$.}
    \KwOut{$\mathcal{D}'$: the sanitized database.}

    \For{$\forall$ $sp_k$ $\in$  \textit{SPIs} }{
         construct the SISL and SIL structures for $sp_k$\;
    }
    sort each itemset in the \textit{SPIs} in descending order based on the number of items they contain\;
    \textit{minSup} = $|\mathcal{D}|$ / \textit{maxAvg} - 1\;
    \For{$\forall$ $sp_k$ $\in$  \textit{SPIs} }{
       get the SISL and SIL of $sp_k$\;
       $du$($sp_k$) = SISL($sp_k$).$SU$ $-$ $\varphi$\;
       $ds$($sp_k$) = SISL($sp_k$).$sup$\;
       $dp$($sp_k$) = SISL($sp_k$).$LP$\;
       \While{$du$($sp_k$) $\ge$ 0 $\&\&$ $dp$($sp_k$) $\le$ \textit{maxPer} $\&\&$ $ds$($sp_k$) $\ge$ \textit{minSup}}{
          
          $T^{select}$ $\gets$ $\max$ \{ $u$($sp_k$, $T_d$), $T_d$ $\in$ SISL($sp_k$).\textit{[tid:utility]} \}\;

          $i^{select}$ $\gets$  $\min$ \{ $mp$($i_d$, $T^{select}$), 1 $\le$ $d$ $\le$ $|sp_k|$, $i_d$ $\in$ SIL($sp_k$, $T^{select}$) \}\;
   
		 \eIf{SIL($sp_k$, $i^{select}$, $T^{select}$).$iu$ $\le$ $du$($sp_k$)}{
			delete $i^{select}$ from $T^{select}$\;
                SISL($sp_k$).$su$ $-$$=$ $u$($i^{select}$, $T^{select}$)\;
                SISL($sp_k$).$sup$ $-$$-$\;
                SISL($sp_k$).$LP$ $=$ \textit{getNewLP}($sp_k$)\;
                update \textit{SPIs}, SISL and SIL\;
		 }{
			$dq$ = $\lceil$ $du$($sp_k$) $/$ $eu$($i^{select}$) $\rceil$\;
                SIL($i^{select}$, $T^{select}$).$cnt$ $-$$=$ $dq$\;
                SIL($i^{select}$, $T^{select}$).$iu$ $-$$=$ $dq$ $\times$ $eu$($i^{select}$)\;
                delete $sp_k$ from \textit{SPIs}\;
                update SISL and SIL\;
		 }
        }
    }
\end{algorithm}

\subsection{Complexity analysis}

The MU-MAP algorithm comprises three main phases: (1) construction of SISL and SIL, (2) sorting of sensitive itemsets, and (3) the hiding phase. Given that there are $m$ sensitive PHUIs in $\mathcal{D}$, with each sensitive itemset containing $k$ items, the construction of each SISL requires $O$($a$) time and space complexity, while building an SIL for each sensitive item demands $O$($b$) time and space complexity. Consequently, the SISL and SIL construction phase exhibits both time and space complexity of $O$($ma$ + $mkb$). During the sensitive itemset sorting phase, the algorithm demonstrates time complexity of $O$($m$$\log$$m$) and space complexity of $O$($m$). In the hiding phase, the outer loop iterates through all sensitive itemsets with $O$($m$) time complexity. Considering $\mathcal{D}$ containing $n$ transactions, the inner while loop may, in the worst-case scenario, need to process every item in each transaction, resulting in $O$($n$) time complexity. For the optimal case, assuming that the sensitive itemset appears in the first transaction, the time complexity is $O$($1$). Each iteration of the inner loop involves transaction and item selection operations with $O$($t$ + $k$) time complexity, yielding an overall time complexity of $O$($m$ $\times$ $n$ $\times$ ($t$ + $k$)) for the hiding phase. In the optimal case, the time complexity of the hiding phase is $O$($m$ $\times$ ($t$ + $k$)). This phase maintains $O$(1) space complexity due to the use of temporary variables for storing victim transactions and items. In summary, the MU-MAP algorithm achieves an overall time complexity of $O$($ma$ + $mkb$ + $m$$\log$$m$ + $m$ $\times$ $n$ $\times$ ($t$ + $k$)) and the space complexity of $O$($ma$ + $mkb$ + $m$ + 1). In the optimal case, the time complexity of the MU-MAP algorithm is $O$($ma$ + $mkb$ + $m$$\log$$m$ + $m$ $\times$ ($t$ + $k$)) and the space complexity is $O$($ma$ + $mkb$ + $m$ + 1). Notice that MU-MIP differs from MU-MAP solely in its victim item selection methodology, thus sharing identical time and space complexity characteristics.

\section{Experiments} \label{sec: experiments}

Previous research has focused on PPUM, yet these studies are inadequate as they fail to consider the periodicity of itemsets. If we apply PPUM algorithms to address the PPPUM problem, several issues arise, including the generation of artificial costs. The LT-MIN and LT-MAX algorithms \cite{gui2024privacy} specifically address privacy preservation for rare itemsets, whereas ODTT \cite{gui2023odtt}, PrivMiner \cite{li2025privminer}, and DP-PartFIT  \cite{liu2024dp} focus on privacy protection in frequent itemset mining. These approaches fundamentally differ from our work in both mining task and technical objectives, rendering them unsuitable for comparative analysis with our proposed method. To illustrate the limitations of PPUM algorithms in this study, we compared our proposed algorithms, MU-MAP and MU-MIP, with the SMAU, SMIU, and SMSE algorithms \cite{liu2020effective}. A series of experiments was conducted on six datasets to evaluate the performance of MU-MAP and MU-MIP against these PPUM algorithms. We expect the experimental results to demonstrate that our proposed algorithms are capable of correctly and effectively addressing the PPPUM problem. Traditional PPUM algorithms, by neglecting the periodicity of itemsets, are likely to produce inaccurate or suboptimal outcomes. All the algorithms in this study were implemented in Java. All experiments were performed on a 64-bit Windows 11 operating system computer, equipped with a 12th Gen Intel Core i7-1260P processor, 16 GB of RAM, and a 2.10 GHz GPU. The code and datasets are publicly available at \href{https://github.com/DSI-Lab1/PPUM}{https://github.com/DSI-Lab1/PPUM.}

\subsection{Datasets and evaluation metrics}

During the experiment, six datasets were tested: kosarak, retail, mushroom, liquor, chainstore, and foodmart. These datasets are real and accessible via the SPMF\footnote{\url{https://www.philippe-fournier-viger.com/spmf/}}. The utility within the datasets was randomly assigned, with internal utility spanning from 10 to 100 and external utility ranging from 1 to 10. Table \ref{CharacteristicsOfDataset} summarizes the key characteristics of these datasets, encompassing (1) the count of transactions (\#T), (2) the count of items (\#I), (3) the average length of transactions (\#TL), and (4) the periodicity thresholds (\#P). Within \#P, the information is presented sequentially as \textit{minPer}, \textit{maxPer}, \textit{minAvg}, and \textit{maxAvg}. 

\begin{itemize}
    \item \textbf{kosarak}: Interactions recorded from click-stream data of a Hungarian news website.

    \item \textbf{retail}: Transactions made by customers at an undisclosed retail outlet in Belgium.

    \item  \textbf{mushroom}: Compiled utilizing the UCI mushrooms dataset as the foundation.

    \item \textbf{liquor}: This dataset consists of 9,284 customer purchase records sourced from liquor stores situated in IOWA.

    \item \textbf{chainstore}: Customer purchases recorded at a prestigious grocery store chain situated in California, USA.

    \item \textbf{foodmart}: Transactions sourced from a retail store, retrieved and reformatted using SQL Server 2000. 
\end{itemize}

\begin{table}[h]
    \centering
    \caption{Characteristics of the dataset}
    \tabcolsep=0.15cm
    \label{CharacteristicsOfDataset}
    \begin{tabular}{ccccc}
        \hline
        \textbf{Dataset} & \textbf{\#T} & \textbf{\#I} & \textbf{\#TL} & \textbf{\#P} \\
        \hline
        kosarak & 990002 & 41270 & 8.10 & 1-2000-1-1000 \\
        retail & 88162 & 16470 & 10.30 & 1-5000-1-2000\\ 
        mushroom & 8416 & 119 & 23.00 & 1-2000-1-1000\\
        liquor & 9284 & 2626 & 2.70 & 1-2000-1-1000 \\
        chainstore & 1112949 & 46086 & 7.23 & 1-6000-1-3000\\ 
        foodmart & 4141 & 1559 & 4.42 & 1-1000-1-500 \\  \hline
    \end{tabular}
\end{table}

Existing methods have proposed various performance evaluation metrics. Our study adopts the most classical performance evaluation criteria, as those in previous studies \cite{wu2006hiding,lin2016fast}. In future work, we may explore superior performance evaluation metrics. Standard evaluation metrics in PPDM include HF, MC, and AC \cite{wu2006hiding}. Since both our algorithms and the SMAU, SMIU, and SMSE algorithms hide SPHUIs by either deleting victim items from victim transactions or reducing the number of victim items, the HF is consistently 0. Therefore, the evaluation focuses solely on the MC and AC of PHUIs discovered in the database before and after the hiding process. We also adopted additional evaluation metrics, including IUS, DUS, and DSS \cite{lin2016fast}. In our study, the SPIs were randomly selected. Specifically, a certain number of itemsets were randomly chosen from all the discovered PHUIs to serve as the sensitive itemsets for hiding. It is important to note that the concept of sensitive percentage (\textit{sep}) is used in this study to control the number of sensitive itemsets. The \textit{sep} refers to the ratio of the selected sensitive itemsets to the total SPIs, which follows a similar approach adopted by previous studies, such as Ref \cite{liu2020effective,gui2024privacy}. The approach to defining dataset periodicity in this work follows a similar methodology as described in previous studies \cite{fournier2016phm,lai2023PHMN}.

\subsection{Artificial cost (AC)}

To illustrate the limitations of traditional PPUM algorithms in addressing the PPPUM problem, we compared the changes in AC with varying \textit{sep} for the MU-MAP, MU-MIP, SMAU, SMIU, and SMSE algorithms across six datasets under the parameter settings specified in this paper. The experimental results are presented in Table \ref{tabAC}. As observed in Table \ref{tabAC}, the proposed MU-MAP and MU-MIP algorithms do not generate any AC when solving the PPPUM problem, whereas the SMAU, SMIU, and SMSE algorithms do produce AC. For instance, in the foodmart dataset, as \textit{sep} increases, the AC generated by the SMAU, SMIU, and SMSE algorithms is 0.18\%, 0.19\%, 0.19\%, 0.76\%, and 0.39\%, respectively. Additionally, in the chainstore, retail, kosarak, mushroom, and liquor datasets, the SMAU, SMIU, and SMSE algorithms also generate AC. In contrast, the MU-MAP and MU-MIP algorithms achieve zero AC across all six datasets. This discrepancy arises because the MU-MAP and MU-MIP algorithms consider the periodicity of itemsets when hiding sensitive periodic information, whereas the SMAU, SMIU, and SMSE algorithms do not account for this aspect. These findings demonstrate that our proposed algorithms can effectively address the PPPUM problem, while the use of PPUM algorithms may introduce AC, causing potential interference for users.

\begin{table*}[htbp]
    \footnotesize
    \centering
    \caption{The AC(\%) under different \textit{sep}.}
\resizebox{\textwidth}{!}{ 
\begin{tabular}{|c|ccccccc|c|ccccccc|}
\hline
\textbf{Dataset} & \textbf{Parameter} & \textbf{test$_1$} & \textbf{test$_2$} & \textbf{test$_3$} & \textbf{test$_4$} & \textbf{test$_5$} & \textbf{test$_6$} & \textbf{Dataset} & \textbf{Parameter} & \textbf{test$_1$} & \textbf{test$_2$} & \textbf{test$_3$} & \textbf{test$_4$} & \textbf{test$_5$} & \textbf{test$_6$}\\ \hline

\multirow{1}{*}{chainstore} & \textit{sep(\%)} & 5 & 6 & 7 & 8 & 9 & 10 & \multirow{1}{*}{foodmart} & \textit{sep(\%)} & 5 & 6 & 7 & 8 & 9 & 10  \\ \cline{2-8} \cline{10-16}
& MU-MAP & 0 & 0 & 0 & 0 & 0 & 0 &  & MU-MAP & 0 & 0 & 0 & 0 & 0 & 0 \\  
& MU-MIP & 0 & 0 & 0 & 0 & 0 & 0 &  & MU-MIP & 0 & 0 & 0 & 0 & 0 & 0\\  
& SMAU & 0.58 & 0.62 & 0 & 0 & 0.61 & 0.63 & & SMAU & 0.18 & 0.18 & 0.19 & 0.19 & 0.76 & 0.39 \\
& SMIU & 0.58 & 0.62 & 0 & 0 & 0.62 & 0.64 & & SMIU & 0.18 & 0.18 & 0.19 & 0.19 & 0.76 & 0.39 \\ 
& SMSE & 0.58 & 0.62 & 0 & 0 & 0.61 & 0.63 & & SMSE & 0.18 & 0.18 & 0.19 & 0.19 & 0.76 & 0.39 \\ \hline

\multirow{1}{*}{retail} & \textit{sep(\%)} & 1 & 2 & 3 & 4 & 5 & 6  & \multirow{1}{*}{kosarak} & \textit{sep(\%)} & 1 & 2 & 3 & 4 & 5 & 6  \\ \cline{2-8} \cline{10-16}
&MU-MAP&0&0 & 0 & 0 & 0 & 0 & & MU-MAP & 0 & 0 & 0 & 0 & 0 & 0\\  
&MU-MIP&0&0 & 0 & 0 & 0 & 0 & & MU-MIP & 0 & 0 & 0 & 0 & 0& 0\\  
&SMAU&0&0& 0 & 0 & 0 & 0 & & SMAU & 0 & 0 & 0 & 0.41 & 0& 0 \\  
&SMIU&0&0& 0 & 0 & 0.08 & 0& & SMIU & 0 & 0 & 0 & 0 & 0& 0 \\  
&SMSE&0&0& 0 & 0 & 0& 0 & & SMSE & 0 & 0 & 0 & 0 & 0 & 0\\ \hline
\multirow{1}{*}{mushroom} & \textit{sep(\%)} & 1 & 2 & 3 & 4 & 5& 6  & \multirow{1}{*}{liquor} & \textit{sep(\%)} & 1 & 2 & 3 & 4 & 5 & 6\\ \cline{2-8} \cline{10-16}
& MU-MAP & 0 & 0 & 0 & 0 & 0& 0 & & MU-MAP & 0 & 0 & 0 & 0 & 0& 0 \\  
& MU-MIP & 0 & 0 & 0 & 0 & 0 & 0& & MU-MIP & 0 & 0 & 0 & 0 & 0& 0\\  
& SMAU & 0 & 0 & 0 & 0 & 0& 0 & & SMAU & 0 & 0 & 0 & 0 & 0 & 0\\  
& SMIU & 0 & 0 & 0 & 0 & 0.08& 0 & & SMIU & 0 & 0 & 0 & 0 & 0 & 0\\  
& SMSE & 0 & 0 & 0 & 0 & 0& 0 & & SMSE & 0 & 0 & 0 & 0 & 0& 0 \\ \hline
\end{tabular}
}
\label{tabAC}
\end{table*}

\subsection{Runtime} \label{experimentRuntime}

\begin{figure*}[h!]
  \centering
  \includegraphics[width=0.9\textwidth]{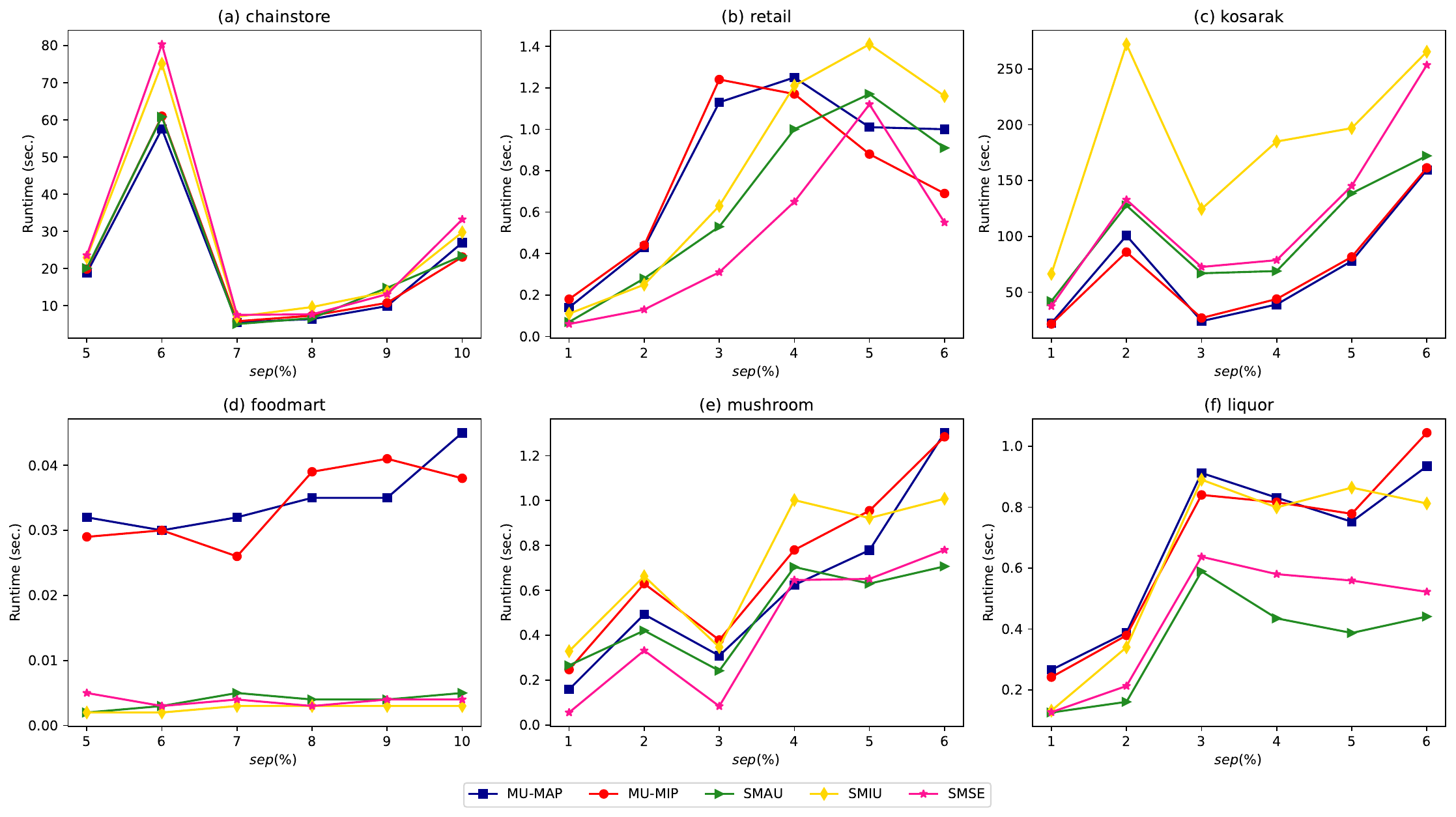} 
  \caption{Runtime under changed \textit{sep}.}
  \label{fig:Runtime}
\end{figure*}

Here, we present the execution times of the MU-MAP and MU-MIP algorithms under different \textit{sep} settings, enabling a clear comparison of their performance. It is important to note that only the execution times for the hiding phase of the algorithms are recorded. The experimental results are illustrated in Fig. \ref{fig:Runtime}. From Fig. \ref{fig:Runtime}, we can observe that the runtime of the MU-MAP, MU-MIP, SMAU, SMIU, and SMSE algorithms generally increases as \textit{sep} grows. This is because a higher \textit{sep} value leads to the selection of more sensitive itemsets, thereby increasing the number of itemsets that need to be hidden. In Fig. \ref{fig:Runtime}(a), we see that when \textit{sep} is 6\%, the execution time of all algorithms is significantly high. However, at \textit{sep} = 7\%, there is a sharp decrease in execution time. This phenomenon occurs because the runtime of the algorithms is closely tied to the number of sensitive itemsets selected. If a sensitive itemset appears in many transactions, hiding these sensitive patterns requires processing numerous transactions. For instance, when \textit{sep} is 6\%, the MU-MAP algorithm updates 125,354 transactions, whereas at \textit{sep} = 7\%, it updates only 45,681 transactions. This reduction in the number of updates by a factor of ten results in a significant decrease in runtime. To validate our analysis, a detailed explanation was provided in Experiment \ref{experimentDiffSelectSPI}. Fig. \ref{fig:Runtime}(b) shows that as \textit{sep} increases, the execution time of the MU-MAP and MU-MIP algorithms initially exceeds that of the SMAU, SMIU, and SMSE algorithms. However, as \textit{sep} continues to grow, the MU-MAP and MU-MIP algorithms outperform the others in terms of runtime. Fig. \ref{fig:Runtime}(c) further illustrates that with increasing \textit{sep}, the runtime of the MU-MAP and MU-MIP algorithms remains consistently lower than that of the SMAU, SMIU, and SMSE algorithms. This improvement is attributed to the consideration of periodicity when selecting victim items in the proposed algorithms, a factor ignored by the SMAU, SMIU, and SMSE algorithms. However, in the foodmart dataset shown in Fig. \ref{fig:Runtime}(d), the runtime of the MU-MAP and MU-MIP algorithms is longer than that of the SMAU, SMIU, and SMSE algorithms. This is because the MU-MAP and MU-MIP algorithms account for the periodicity of itemsets during the sensitive information hiding process. Calculating and recording the periodicity of sensitive itemsets requires additional computation time, which the SMAU, SMIU, and SMSE algorithms do not incur, as they disregard periodicity information.

In some datasets, runtime exhibits a fluctuating upward trend. For instance, in Fig. \ref{fig:Runtime}(e), when \textit{sep} = 0.3, the runtime of the MU-MAP, MU-MIP, SMAU, SMIU, and SMSE algorithms decreases. This behavior is due to the random selection of sensitive itemsets from the SPIs. If the selected itemsets do not overlap—meaning the itemsets are not present in the same transactions or share a few common items—the algorithms must process each itemset almost independently. However, if there is overlap among the items in the selected sensitive itemsets, processing overlapping items in one itemset affects the handling of other sensitive itemsets, reducing the total number of processing iterations and thereby lowering runtime. Overall, the execution time of MU-MAP and MU-MIP demonstrates a relatively better performance trend.

\subsection{Missing cost (MC)}

\begin{figure*}[h!]
  \centering
  \includegraphics[width=0.9\textwidth]{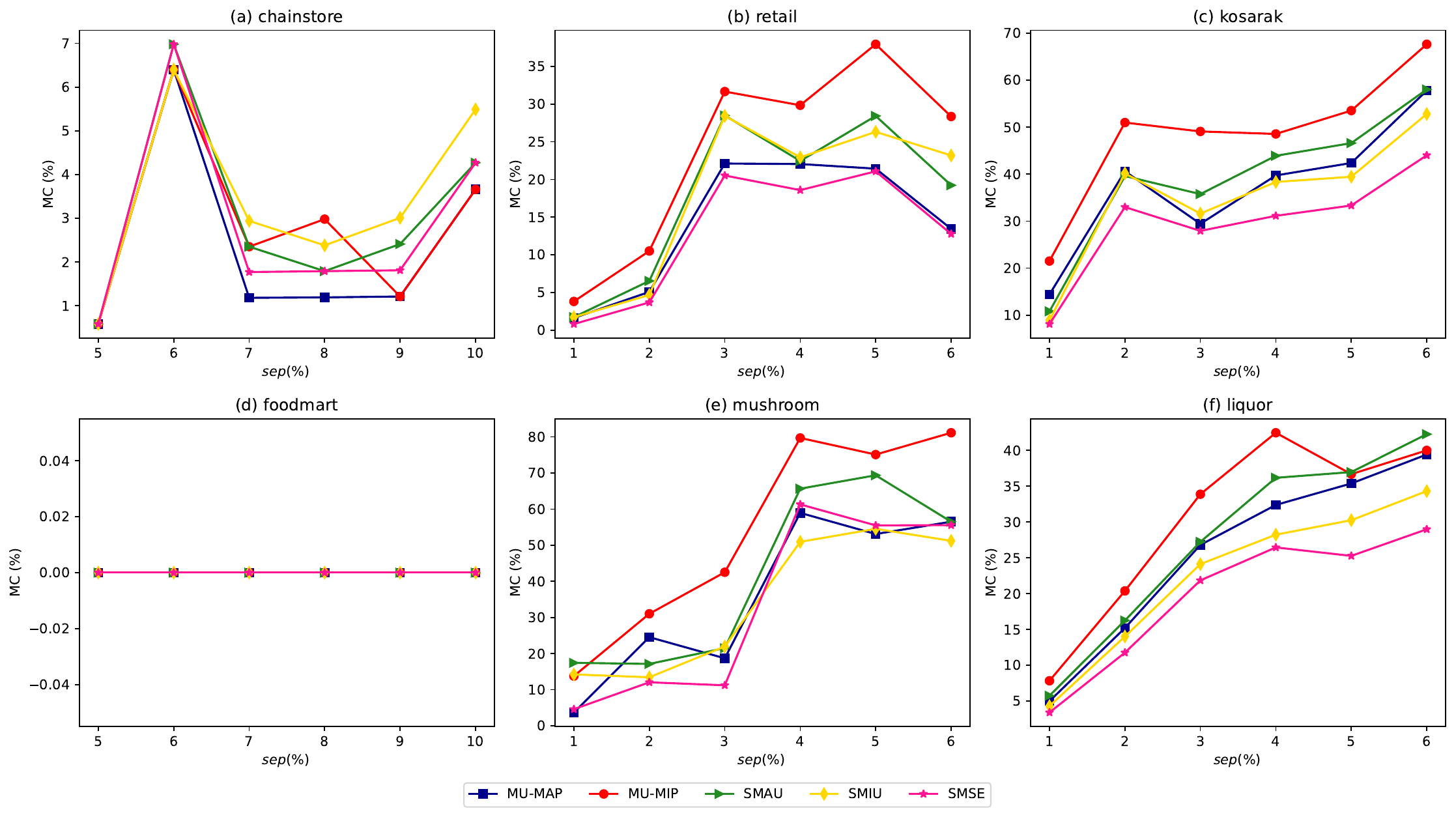} 
  \caption{MC under changed \textit{sep}.}
  \label{fig:MC}
\end{figure*}

To further compare the performance of the MU-MAP and MU-MIP algorithms, we evaluated the MC of the MU-MAP, MU-MIP, SMAU, SMIU, and SMSE algorithms across six different datasets. The experimental results are presented in Fig. \ref{fig:MC}. From Fig. \ref{fig:MC}, we can observe that all algorithms achieve an MC of 0 on the foodmart dataset. This is because foodmart is a sparse dataset where the selected sensitive itemsets rarely share common items, meaning that hiding these sensitive itemsets does not affect other itemsets. Additionally, the number of periodic itemsets in foodmart is relatively small, leading to fewer sensitive itemsets being selected. In Fig. \ref{fig:MC}(a), when \textit{sep} = 6\%, the MC for all algorithms is relatively high, but it decreases significantly at \textit{sep} = 7\%. This trend occurs because the MC is closely tied to the selection of sensitive itemsets. If sensitive itemsets and non-sensitive itemsets share common items and co-occur in the same transaction, hiding sensitive itemsets can inadvertently affect non-sensitive itemsets. Consequently, some itemsets may be missing when mining PHUIs from the sanitized dataset. As shown in Fig. \ref{fig:MC}, the MC of MU-MIP is higher than that of the other algorithms, while MU-MAP demonstrates a comparatively lower MC. The SMSE algorithm exhibits a lower MC than the others on the retail, kosarak, and liquor datasets. This discrepancy arises because the MU-MIP and MU-MAP algorithms prioritize the utility value of sensitive itemsets when selecting victim transactions. In contrast, the SMAU, SMIU, and SMSE algorithms focus on transactions containing the fewest non-sensitive itemsets. This selection strategy allows SMAU, SMIU, and SMSE to minimize the impact on non-sensitive information during the sensitive information-hiding process. Furthermore, MU-MAP prioritizes victim items with the highest periodicity for removal or reduction. Removing such items increases the periodicity of the itemsets, making it easier to exceed the periodicity constraints and more effectively hide sensitive itemsets, thus reducing the number of missed itemsets. Besides, the MU-MIP algorithm prioritizes victim items with the smallest periodicity, which results in slower increases in periodicity when these items are reduced or removed. This necessitates multiple iterations to process sensitive itemsets, thereby increasing the number of missed itemsets. In contrast, the SMAU, SMIU, and SMSE algorithms do not consider the periodicity information of itemsets when selecting victim items. Overall, the MU-MAP algorithm demonstrates relatively low MC when addressing the PPPUM problem, highlighting its effectiveness in minimizing missed itemsets.

\subsection{Itemset utility similarity (IUS)}

Here, the IUS metric \cite{lin2016fast} is used to evaluate the utility loss of PHUIs found in the dataset before and after sanitization. IUS is defined as follows:
$$
\textit{IUS} = \frac{\sum_{Y\in \textit{PHUIs}^{D'}}{u\left( Y \right)}}{\sum_{Y\in \textit{PHUIs}^D}{u\left( Y \right)}}
$$
where \textit{PHUIs}$^D$ denotes the PHUIs discovered from the original dataset $\mathcal{D}$, and \textit{PHUIs}$^{D'}$ represents the PHUIs discovered from the sanitized dataset $\mathcal{D}'$.

We evaluated the IUS of the MU-MAP, MU-MIP, SMAU, SMIU, and SMSE algorithms under different \textit{sep} settings across six real-world datasets. The experimental results are presented in Table \ref{tabIUS}. From Table \ref{tabIUS}, it can be observed that on the foodmart dataset, the IUS of the MU-MAP and MU-MIP algorithms reached 100\%, while the IUS of the SMAU, SMIU, and SMSE algorithms exceeded 100\%. This phenomenon occurs because the MU-MAP and MU-MIP algorithms correctly solve the PPPUM problem, whereas the SMAU, SMIU, and SMSE algorithms, as indicated in Table \ref{tabAC}, generate AC due to their disregard for the periodicity of itemsets. As shown in Fig. \ref{fig:MC}, the MC of all five algorithms is 0 on the foodmart dataset. However, the AC introduced by SMAU, SMIU, and SMSE leads to the discovery of HUIs in the sanitized dataset that do not exist in the original dataset, causing their IUS to exceed 100\%. On the chainstore and mushroom datasets, the IUS of the MU-MAP and MU-MIP algorithms is higher than that of the SMAU, SMIU, and SMSE algorithms. This is because the MU-MAP and MU-MIP algorithms account for the periodicity information of itemsets when selecting victim items, whereas SMAU, SMIU, and SMSE do not. On the retail, kosarak, mushroom, and liquor datasets, the MU-MIP algorithm exhibits the lowest IUS, while the IUS of the MU-MAP algorithm closely matches that of SMSE, the best-performing algorithm among the PPUM algorithms. This difference arises from the strategies used for selecting victim items. The MU-MAP algorithm prioritizes items with the highest periodicity as victim items. Reducing or removing such items quickly increases the periodicity of sensitive itemsets, allowing them to exceed periodic constraints and thereby achieve effective hiding. In contrast, the MU-MIP algorithm prioritizes items with the lowest periodicity, resulting in slower increases in periodicity for sensitive itemsets. Consequently, more itemsets must be modified, leading to greater utility reduction and a relatively lower IUS. Both MU-MAP and MU-MIP consider the utility values of itemsets when selecting victim transactions, whereas the SMAU, SMIU, and SMSE algorithms prioritize the number of non-sensitive itemsets within transactions. On the foodmart and chainstore datasets, the IUS of all algorithms is similar. These datasets are sparse, with fewer items in the selected sensitive itemsets and shorter transaction lengths for transactions containing sensitive itemsets. As a result, the number of transactions processed by the five algorithms is approximately the same, leading to a similar impact on PHUIs and comparable IUS values. Additionally, the overlap of sensitive items within sensitive itemsets also influences algorithm performance. If a sensitive item appears in multiple sensitive itemsets, processing that item affects several sensitive itemsets simultaneously, reducing the number of operations required. Overall, the proposed algorithms demonstrate strong performance, achieving favorable results across various datasets.

\begin{table*}[htbp]
    \footnotesize
    \centering
    \caption{The IUS under different \textit{sep}.}
\resizebox{\textwidth}{!}{ 
\begin{tabular}{|c|ccccccc|c|ccccccc|}
\hline
\textbf{Dataset} & \textbf{Parameter} & \textbf{test$_1$} & \textbf{test$_2$} & \textbf{test$_3$} & \textbf{test$_4$} & \textbf{test$_5$} & \textbf{test$_6$} & \textbf{Dataset} & \textbf{Parameter} & \textbf{test$_1$} & \textbf{test$_2$} & \textbf{test$_3$} & \textbf{test$_4$} & \textbf{test$_5$} & \textbf{test$_6$}\\ \hline

\multirow{1}{*}{chainstore} & \textit{sep(\%)} &5 & 6 & 7 & 8 & 9 & 10 & \multirow{1}{*}{foodmart} & \textit{sep(\%)}&5 & 6 & 7 & 8 & 9 & 10  \\ \cline{2-8} \cline{10-16}
&MU-MAP&99.83& 93.32 & 98.67 & 98.74 & 99.13 & 96.99 &&MU-MAP&100.00 & 100.00 & 100.00 & 100.00 & 100.00 & 100.00 \\  
&MU-MIP&99.83& 93.32 & 99.10 & 97.99 & 99.39 & 96.40 &&MU-MIP& 100.00 &100.00 & 100.00 & 100.00 & 100.00 & 100.00\\  
&SMAU&100.02& 93.05 & 97.55 & 98.35 & 97.22 & 95.62 &&SMAU& 100.08 &100.08 & 100.08 & 100.08 & 100.43 & 100.17 \\
&SMIU&100.02& 93.41 & 98.79 & 98.13 & 97.94 & 96.37 &&SMIU&100.08 & 100.08 & 100.08 & 100.08 & 100.43 & 100.17 \\ 
&SMSE&100.02& 93.10 & 98.10 & 98.44 & 98.31 & 95.73 &&SMSE&100.08 & 100.08 & 100.08 & 100.08 & 100.43 & 100.17 \\ \hline

\multirow{1}{*}{retail} & \textit{sep(\%)} & 1 & 2 & 3 & 4 & 5& 6  & \multirow{1}{*}{kosarak} & \textit{sep(\%)} & 1 & 2 & 3 & 4 & 5& 6  \\ \cline{2-8} \cline{10-16}
& MU-MAP & 98.39 & 95.09 & 75.81 & 74.25 & 76.13& 83.29 &&MU-MAP& 87.45 & 60.36 & 73.93 & 60.95 & 63.59&46.16 \\  
& MU-MIP & 95.82 & 88.58 & 67.23 & 67.81 & 61.63&70.53 &&MU-MIP& 78.50 & 52.40 & 54.55 & 54.46 & 50.43&37.56\\  
& SMAU & 97.89 & 92.50 & 69.07 & 73.31 & 68.30&76.50 &&SMAU& 87.13 & 60.20 & 64.64 & 57.34 & 55.56&43.57 \\  
& SMIU & 98.24 & 94.82 & 71.05 & 75.68 & 74.26&75.52 &&SMIU& 90.50 & 61.17 & 70.24 & 64.02 & 62.97&49.90 \\  
& SMSE & 98.87 & 95.80 & 78.27 & 78.31 & 77.05&83.96 &&SMSE& 91.48 & 65.25 & 72.90 & 70.75 & 68.83 &56.00\\ \hline

\multirow{1}{*}{mushroom} & \textit{sep(\%)} & 1 & 2 & 3 & 4 & 5&6  & \multirow{1}{*}{liquor} & \textit{sep(\%)} & 1 & 2 & 3 & 4 & 5&6  \\ \cline{2-8} \cline{10-16} 
& MU-MAP & 96.48 & 74.74 & 79.65 & 40.34 & 46.65&41.72 &&MU-MAP& 95.00 & 85.80 & 74.96 & 69.66 & 66.62&62.67 \\  
& MU-MIP & 83.30 & 65.72 & 55.21 & 19.20 & 23.51&17.59 &&MU-MIP& 90.59 & 78.81 & 65.69 & 56.20 & 61.35&58.52\\  
& SMAU & 82.59 & 81.68 & 77.62 & 32.54 & 29.19&41.35 &&SMAU& 93.31 & 82.68 & 72.77 & 62.26 & 61.52&55.84 \\  
& SMIU & 85.85 & 85.50 & 77.70 & 47.50 & 43.12&46.86 &&SMIU& 95.84 & 86.36 & 76.39 & 72.08 & 69.70&65.72 \\  
& SMSE & 95.49 & 86.32 & 88.41 & 38.44 & 43.24&43.51 &&SMSE& 96.38 & 88.22 & 78.66 & 74.17 & 73.97 &71.03\\ \hline
\end{tabular}
}
\label{tabIUS}
\end{table*}

\subsection{Database utility similarity (DUS)}

Here, DUS \cite{lin2016fast} is adopted to assess the similarity in utility between the original and sanitized datasets. The DUS metric is defined as follows:
$$
\textit{DUS} = \frac{\sum_{T_u\in D'}{tu\left( T_u \right)}}{\sum_{T_u\in D}{tu\left( T_u \right)}}
$$
where $tu$($T_u$) represents the utility of the transaction $T_u$.

Table \ref{tabDUS} presents the DUS of the MU-MAP, MU-MIP, SMAU, SMIU, and SMSE algorithms under different \textit{sep} settings. From Table \ref{tabDUS}, it is evident that the MU-MAP and MU-MIP algorithms achieve higher DUS values than the SMAU, SMIU, and SMSE algorithms on the chainstore and foodmart datasets. This is because the MU-MAP and MU-MIP algorithms consider the periodicity information of itemsets, while the SMAU, SMIU, and SMSE algorithms do not. By incorporating periodicity constraints in addition to utility threshold constraints, the MU-MAP and MU-MIP algorithms can more effectively hide sensitive itemsets, thereby minimizing the utility loss in the database. For the other datasets, the DUS values of the MU-MAP, MU-MIP, SMAU, SMIU, and SMSE algorithms are similar. This occurs because all algorithms use the same sensitive itemsets for sanitization, which necessitates a loss of utility value of these sensitive itemsets in the database. However, different strategies for selecting victim items influence the impact on non-sensitive itemsets, resulting in varying utility losses. The MU-MAP and MU-MIP algorithms select victim items based on their maximum periodicity, while the SMAU and SMIU algorithms prioritize items based on utility values. The SMSE algorithm, in contrast, selects items based on the number of sensitive and non-sensitive itemsets affected by the victim items.

In summary, the MU-MAP and MU-MIP algorithms are effective in reducing database utility loss while hiding sensitive itemsets. The DUS of the sanitized database relative to the original dataset remains above 94\%. This high score is attributed to the fact that the transactions containing sensitive itemsets and the utility values of the sensitive itemsets remain fixed. Regardless of the hiding strategy, the algorithms only need to reduce or remove victim items from the transactions containing sensitive itemsets to alter their utility or periodicity, thus achieving the objective of hiding. The minimal changes to victim items in periodic high-utility itemsets during the sanitization process contribute to the high DUS values, underscoring the efficiency and effectiveness of the proposed algorithms.

\begin{table*}[htbp]
    \footnotesize
    \centering
    \caption{The DUS under different \textit{sep}.}
\resizebox{\textwidth}{!}{ 
\begin{tabular}{|c|ccccccc|c|ccccccc|}
\hline
\textbf{Dataset} & \textbf{Parameter} & \textbf{test$_1$} & \textbf{test$_2$} & \textbf{test$_3$} & \textbf{test$_4$} & \textbf{test$_5$}& \textbf{test$_6$} & \textbf{Dataset} & \textbf{Parameter} & \textbf{test$_1$} & \textbf{test$_2$} & \textbf{test$_3$} & \textbf{test$_4$} & \textbf{test$_5$}& \textbf{test$_6$}\\ \hline

\multirow{1}{*}{chainstore} & \textit{sep(\%)}&5 & 6 & 7 & 8 & 9 & 10 & \multirow{1}{*}{foodmart} & \textit{sep(\%)} &5& 6 & 7 & 8 & 9 & 10  \\ \cline{2-8} \cline{10-16}
&MU-MAP&99.30& 98.30 & 99.03 & 99.35 & 99.35 & 98.11 &&MU-MAP&98.88& 99.00 & 98.35 & 98.20 & 97.92 & 97.74 \\  
&MU-MIP&99.30& 98.30 & 99.09 & 99.34 & 99.37 & 98.09 &&MU-MIP&98.88& 99.00 & 98.35 & 98.20 & 97.92 & 97.74\\  
&SMAU&99.14& 98.21 & 98.74 & 99.01 & 98.85 & 97.57 &&SMAU&98.46& 98.65 & 97.86 & 97.45 & 97.10 & 96.82 \\
&SMIU&99.14& 98.23 & 98.82 & 99.03 & 98.93 & 97.60 &&SMIU&98.46& 98.65 & 97.86 & 97.45 & 97.10 & 96.82 \\ 
&SMSE&99.14& 98.22 & 98.75 & 99.02 & 98.91 & 97.58 &&SMSE&98.46& 98.65 & 97.86 & 97.45 & 97.10 & 96.82 \\ \hline

\multirow{1}{*}{retail} & \textit{sep(\%)} & 1 & 2 & 3 & 4 & 5 &6 & \multirow{1}{*}{kosarak} & \textit{sep(\%)} & 1 & 2 & 3 & 4 & 5&6  \\ \cline{2-8} \cline{10-16} 
& MU-MAP & 99.78 & 99.29 & 97.35 & 96.62 & 96.94&97.86 &&MU-MAP& 99.28 & 97.89 & 99.01 & 98.32 & 97.87&96.32\\ 
& MU-MIP & 99.84 & 99.22 & 98.14 & 97.90 & 98.00&98.13 &&MU-MIP& 99.20 & 99.31 & 99.32 & 99.06 & 99.05&97.88\\ 
& SMAU & 99.69 & 99.02 & 96.94 & 96.34 & 96.45&97.12 &&SMAU& 99.01 & 97.79 & 98.20 & 97.71 & 97.10 &95.62\\  
& SMIU & 99.91 & 99.44 & 98.59 & 98.28 & 98.87&98.49 &&SMIU& 99.32 & 99.33 & 99.45 & 99.23 & 99.15 &98.18\\  
& SMSE & 99.80 & 99.28 & 97.76 & 96.91 & 97.75&97.98 &&SMSE& 99.28 & 98.56 & 99.21 & 98.83 & 98.56&97.32 \\ \hline

\multirow{1}{*}{mushroom} & \textit{sep(\%)} & 1 & 2 & 3 & 4 & 5&6  & \multirow{1}{*}{liquor} & \textit{sep(\%)} & 1 & 2 & 3 & 4 & 5&6  \\ \cline{2-8} \cline{10-16}
& MU-MAP & 99.71 & 98.09 & 99.31 & 98.30 & 97.37&97.65 &&MU-MAP& 99.27 & 98.05 & 95.40 & 94.95 & 94.64&93.50\\ 
& MU-MIP & 99.77 & 99.10 & 99.49 & 98.79 & 98.79&98.27 &&MU-MIP& 99.06 & 97.93 & 94.89 & 94.04 & 95.02&94.00\\ 
& SMAU & 99.36 & 97.81 & 99.16 & 97.61 & 96.92&96.13 &&SMAU& 98.87 & 97.27 & 94.61 & 92.86 & 93.02 &91.44\\  
& SMIU & 99.87 & 99.43 & 99.88 & 99.40 & 99.41&98.83 &&SMIU& 99.59 & 98.63 & 96.07 & 96.06 & 96.18 &95.12\\  
& SMSE & 99.71 & 98.95 & 99.44 & 98.89 & 98.63&98.43 &&SMSE& 99.40 & 98.14 & 95.72 & 95.37 & 95.42 &94.49\\ \hline
\end{tabular}
}
\label{tabDUS}
\end{table*}

\subsection{Database structure similarity (DSS)}

The DSS \cite{lin2016fast} is also utilized to measure the structural similarity between the original and sanitized datasets. The DSS metric is defined as follows:
$$
\textit{DSS} = \frac{\sum_{k=1}^{\max \left\{ \left| D \right|,\ \left| D' \right| \right\}}{f\left( t_{k}^{D} \right) \cdot f\left( t_{k}^{D'} \right)}}{\sqrt{\sum_{k=1}^{\left| D \right|}{f^2\left( t_{k}^{D} \right)}}\cdot \sqrt{\sum_{k=1}^{\left| D' \right|}{f^2\left( t_{k}^{D'} \right)}}}
$$
where $f$($t_{k}^{D}$) represents the frequency of transaction $t_{k}$ in the original dataset $\mathcal{D}$, and $f$($t_{k}^{D'}$) represents its frequency in the sanitized dataset $\mathcal{D}'$.

Table \ref{tabDSS} records the DSS results for the MU-MAP, MU-MIP, SMAU, SMIU, and SMSE algorithms under various \textit{sep} settings. From the table, it is evident that for the datasets foodmart, retail, kosarak, mushroom, and liquor, the DSS values of MU-MAP and MU-MIP surpass those of SMAU, SMIU, and SMSE. However, in the chainstore dataset, MU-MAP and MU-MIP exhibit lower DSS values compared to SMAU, SMIU, and SMSE. This difference stems from the fact that MU-MAP and MU-MIP incorporate periodicity information when hiding sensitive itemsets, reducing the number of transactions that need modification. In contrast, chainstore, being a sparse dataset with fewer itemsets per transaction, shows SMAU, SMIU, and SMSE algorithms prioritizing victim transactions containing the largest number of non-sensitive itemsets. The data in Table \ref{tabDSS} further reveals that MU-MAP and MU-MIP are effective in minimizing database changes while hiding sensitive itemsets. In datasets such as retail, kosarak, chainstore, and foodmart, both algorithms achieve DSS values exceeding 95\%. For the liquor dataset, the DSS values remain robust, reaching approximately 85\%. However, in the mushroom dataset, the DSS values for MU-MAP and MU-MIP show a slight decline. This is due to the fixed set of transactions containing sensitive itemsets. The algorithms reduce or remove victim items from these transactions to lower the utility or increase the periodicity of sensitive itemsets, thereby achieving the goal of hiding them with minimal modifications. In dense datasets like mushroom, where many transactions contain specific itemsets, the utility increases and periodicity decreases. Consequently, more transactions must be modified to hide sensitive itemsets, leading to a slight reduction in DSS.

Overall, the proposed algorithms demonstrate excellent performance in preserving the database's structure while effectively hiding sensitive itemsets. These results highlight the algorithms' efficiency and effectiveness in solving the PPPUM problem with minimal dataset disruption.

\begin{table*}[htbp]
    \footnotesize
    \centering
    \caption{The DSS under different \textit{sep}.}
\resizebox{\textwidth}{!}{ 
\begin{tabular}{|c|ccccccc|c|ccccccc|}
\hline
\textbf{Dataset} & \textbf{Parameter} & \textbf{test$_1$} & \textbf{test$_2$} & \textbf{test$_3$} & \textbf{test$_4$} & \textbf{test$_5$} & \textbf{test$_6$}& \textbf{Dataset} & \textbf{Parameter} & \textbf{test$_1$} & \textbf{test$_2$} & \textbf{test$_3$} & \textbf{test$_4$} & \textbf{test$_5$}& \textbf{test$_6$}\\ \hline

\multirow{1}{*}{chainstore} & \textit{sep(\%)} &5& 6 & 7 & 8 & 9 & 10 & \multirow{1}{*}{foodmart} & \textit{sep(\%)} &5& 6 & 7 & 8 & 9 & 10  \\ \cline{2-8} \cline{10-16}
&MU-MAP&97.74& 96.25 & 99.44 & 99.92 & 98.93 & 98.15 &&MU-MAP&97.39& 97.98 & 96.29 & 96.45 & 95.75 & 95.56 \\  
&MU-MIP&97.74& 96.25 & 99.44 & 99.92 & 98.93 & 98.32 &&MU-MIP&97.39& 97.98 & 96.29 & 96.45 & 95.75 & 95.56\\  
&SMAU&99.12& 98.91 & 99.70 & 99.90 & 99.53 & 99.46 &&SMAU&95.33& 95.99 & 93.80 & 93.06 & 91.80 & 91.18 \\
&SMIU&99.12& 98.91 & 99.70 & 99.90 & 99.53 & 99.46 &&SMIU&95.33& 95.99 & 93.80 & 93.06 & 91.80 & 91.18 \\ 
&SMSE&99.12& 98.91 & 99.70 & 99.90 & 99.53 & 99.46 &&SMSE&95.33& 95.99 & 93.80 & 93.06 & 91.80 & 91.18 \\ \hline

\multirow{1}{*}{retail} & \textit{sep(\%)} & 1 & 2 & 3 & 4 & 5&6  & \multirow{1}{*}{kosarak} & \textit{sep(\%)} & 1 & 2 & 3 & 4 & 5 &6 \\ \cline{2-8} \cline{10-16} 
& MU-MAP & 99.90 & 99.60 & 98.44 & 97.83 & 98.53&98.71 &&MU-MAP& 99.99 & 99.98 & 99.99 & 99.98 & 99.98&99.95\\ 
& MU-MIP & 99.90 & 99.61 & 98.06 & 97.93 & 98.63&98.81 &&MU-MIP& 99.99 & 99.98 & 99.99 & 99.98 & 99.98&99.94\\ 
& SMAU & 99.85 & 99.46 & 98.20 & 97.36 & 97.99&98.37 &&SMAU& 99.97 & 99.97 & 99.97 & 99.88 & 99.91 &99.86\\  
& SMIU & 99.85 & 99.46 & 97.97 & 97.38 & 98.01&98.32 &&SMIU& 99.97 & 99.97 & 99.97 & 99.88 & 99.92&99.84 \\  
& SMSE & 99.85 & 99.46 & 98.23 & 97.40 & 98.13&98.36 &&SMSE& 99.97 & 99.96 & 99.97 & 99.88 & 99.91 &99.83\\ \hline

\multirow{1}{*}{mushroom} & \textit{sep(\%)} & 1 & 2 & 3 & 4 & 5 &6 & \multirow{1}{*}{liquor} & \textit{sep(\%)} & 1 & 2 & 3 & 4 & 5 &6 \\ \cline{2-8} \cline{10-16} 
& MU-MAP & 95.57 & 78.57 & 93.63 & 82.75 & 77.72&72.55 &&MU-MAP& 97.48 & 94.38 & 88.01 & 86.23 & 88.18&85.44\\ 
& MU-MIP & 96.64 & 85.35 & 94.20 & 87.39 & 81.83&75.33 &&MU-MIP& 97.46 & 94.41 & 87.87 & 86.42 & 88.36&85.61\\ 
& SMAU & 93.10 & 79.02 & 91.01 & 75.72 & 71.41&58.99 &&SMAU& 96.49 & 92.18 & 84.98 & 82.14 & 83.69&80.02 \\  
& SMIU & 93.19 & 78.93 & 90.88 & 75.66 & 71.23&58.84 &&SMIU& 96.48 & 92.15 & 84.99 & 82.13 & 83.56&79.93 \\  
& SMSE & 92.88 & 80.33 & 89.31 & 77.66 & 74.01&62.20 &&SMSE& 96.63 & 92.41 & 85.63 & 82.49 & 84.65&80.91 \\ \hline
\end{tabular}
}
\label{tabDSS}
\end{table*}

\subsection{Analysis of sensitive itemsets(SPIs) selection strategy} \label{experimentDiffSelectSPI}

\begin{figure*}[h!]
  \centering
  \includegraphics[width=0.9\textwidth]{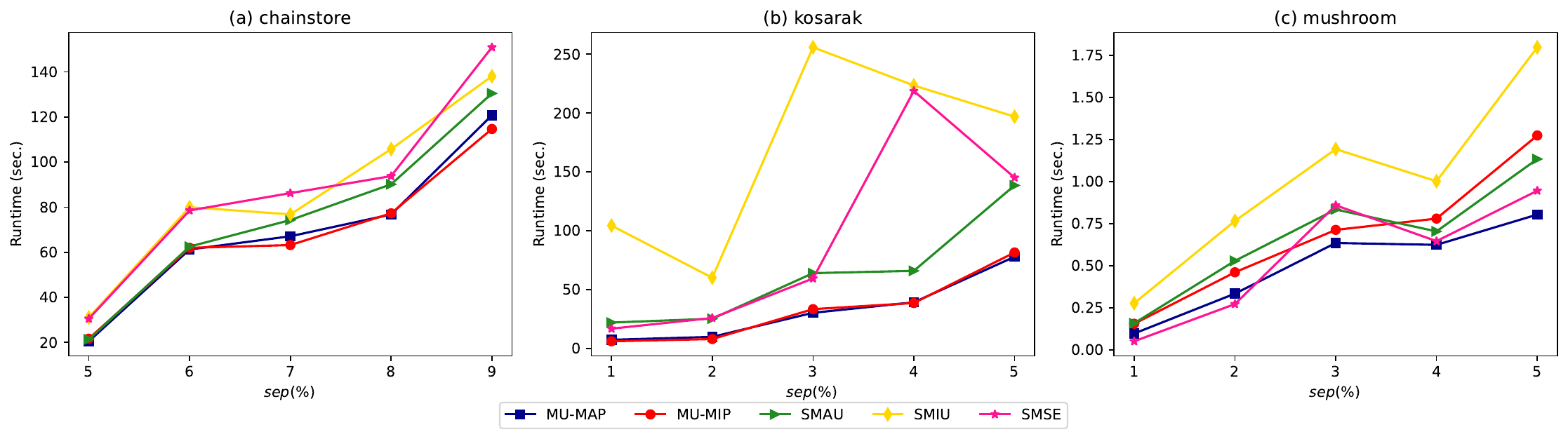} 
  \caption{Runtime under changed \textit{sep}.}
  \label{fig:diffselectspiruntime}
\end{figure*}

To validate the analysis presented in Experiment \ref{experimentRuntime}—that the algorithm's execution time is influenced by the selection of sensitive itemsets—we conducted a series of experiments. As the value of \textit{sep} increases, the number of selected sensitive itemsets also increases. The selection strategy was designed to ensure fairness: sensitive itemsets were incrementally added based on the previously selected ones, thereby maintaining maximum consistency across experiments. All other parameters were kept the same as in Experiment \ref{experimentRuntime}. Due to space limitations, we focused on evaluating execution time using only the kosarak, chainstore, and mushroom datasets to test our hypothesis. The experimental results are shown in Fig. \ref{fig:diffselectspiruntime}. As observed in Fig. \ref{fig:diffselectspiruntime}, with the increase of \textit{sep}, the execution times of both MU-MAP and MU-MIP generally show an upward trend. This strongly supports the analysis in Experiment \ref{experimentRuntime}, confirming that the algorithm's runtime is indeed related to the choice of sensitive itemsets. Moreover, MU-MAP and MU-MIP consistently outperform SMAU, SMIU, and SMSE in terms of execution time. This is because MU-MAP and MU-MIP take the periodicity of itemsets into account when hiding sensitive patterns, whereas SMAU, SMIU, and SMSE only consider the utility values of itemsets. Additionally, the method used for selecting sensitive itemsets also impacts algorithm performance. On the kosarak dataset, the execution times of SMAU, SMIU, and SMSE exhibit noticeable fluctuations. This is because these algorithms are influenced not only by the characteristics and distribution of the sensitive itemsets within the dataset but also by the specific strategies they employ to hide them. In conclusion, the experimental results confirm that the execution time of the algorithms is closely related to the selection of sensitive itemsets.

\subsection{Significance analysis}

In this section, we conducted a two-way analysis of variance (ANOVA) to investigate the effects of different parameters on runtime performance. The experimental evaluation was performed across six distinct datasets, with each experimental configuration replicated three times to ensure statistical reliability. The significance threshold was established at 0.05 for all tests. The F-value represents the ratio of between-group variance to within-group variance (mean square), and the P-value determines whether the observed differences are statistically significant. P $<$ 0.05 indicates a significant difference; P $\geq$ 0.05 indicates no significant difference. In our analysis, $@$\textit{sep} denotes the effect of the \textit{sep} on results, $@$\textit{algorithm} represents the effect of different algorithms, and $@$\textit{sep}$*$\textit{algorithm} indicates their interaction effect. Table \ref{tabSA} presents the two-way ANOVA results with "algorithm type" and "number of SPIs" as independent factors. The analysis reveals three key findings: \textbf{Main effect of algorithm:} Significant differences in runtime were observed across algorithms. For example, all P-values for the $@$\textit{algorithm} were $<$ 0.05 across tested datasets in Table \ref{tabSA}, confirming that algorithm type significantly affects runtime. \textbf{Main effect of \textit{sep} value:} The \textit{sep} significantly influenced runtime performance. For instance, all P-values for $@$\textit{sep} were $<$ 0.05 in Table \ref{tabSA}, demonstrating \textit{sep}'s substantial impact on performance. \textbf{Interaction effect:} The algorithm and \textit{sep} interaction showed statistically significant effects on runtime. For example, all P-values for $@$\textit{sep}$*$\textit{algorithm} were $<$ 0.05 in Table \ref{tabSA}, indicating combined influences on performance.

\begin{table*}[htbp]
    \footnotesize
    \centering
    \caption{The significance analysis of runtime under different \textit{sep} and \textit{algorithms}.}
\resizebox{\textwidth}{!}{ 
\begin{tabular}{|c|ccc|c|ccc|c|ccc|}
\hline
\textbf{Dataset} & \textbf{test} & \textbf{F} & \textbf{P} & \textbf{Dataset} & \textbf{test} & \textbf{F} & \textbf{P} & \textbf{Dataset} & \textbf{test} & \textbf{F} & \textbf{P}\\ \hline

\multirow{1}{*}{chainstore} & \textit{$@$sep}&1554.04 & $<$0.0001 & 
\multirow{1}{*}{foodmart} & \textit{$@$sep} &7.80& $<$0.0001 &   
\multirow{1}{*}{kosarak} & \textit{$@$sep} &193.70& $<$0.0001
\\ \cline{2-4} \cline{6-8} \cline{10-12}
&$@$\textit{algorithm}&35.45& $<$0.0001 &&$@$\textit{algorithm}&521.69& $<$0.0001 &&$@$\textit{algorithm}&142.92& $<$0.0001 \\ \cline{2-4} \cline{6-8} \cline{10-12}
&$@$\textit{sep}$*$\textit{algorithm}&8.71& $<$0.0001 &&$@$\textit{sep}$*$\textit{algorithm}&3.44& 0.0001 &&$@$\textit{sep}$*$\textit{algorithm}&6.10& $<$0.0001 \\
 \hline

\multirow{1}{*}{liquor} & \textit{$@$sep}&222.60 & $<$0.0001 & 
\multirow{1}{*}{mushroom} & \textit{$@$sep} &666.77& $<$0.0001 &   
\multirow{1}{*}{retial} & \textit{$@$sep} &41.36& $<$0.0001
\\ \cline{2-4} \cline{6-8} \cline{10-12}
&$@$\textit{algorithm}&94.09& $<$0.0001 &&$@$\textit{algorithm}&133.69& $<$0.0001 &&$@$\textit{algorithm}&6.25& 0.0003 \\ \cline{2-4} \cline{6-8} \cline{10-12}
&$@$\textit{sep}$*$\textit{algorithm}&4.94& $<$0.0001 &&$@$\textit{sep}$*$\textit{algorithm}&17.78& $<$0.0001 &&$@$\textit{sep}$*$\textit{algorithm}&2.14& 0.0123 \\
 \hline

\end{tabular}
}
\label{tabSA}
\end{table*}

\subsection{Sensitive analysis}

This subsection presents sensitivity analysis experiments to evaluate the impact of different parameter settings on the proposed algorithms. Specifically, the experiments examine how varying the parameters \textit{minutil} and \textit{maxPer} affects the number of selected sensitive itemsets, execution time, and MC of the MU-MAP and MU-MIP algorithms. Through this analysis, we aim to assess the sensitivity of the results to these parameters. Three datasets were selected for the experiments: chainstore, kosarak, and mushroom. For these datasets, all periodicity-related parameters, except for \textit{maxPer}, are set as shown in Table \ref{CharacteristicsOfDataset}. To eliminate the influence of sensitive itemset selection strategies, the experiments adopt an incremental selection approach, where sensitive itemsets are added cumulatively.

\begin{figure*}[h!]
  \centering
  \includegraphics[width=0.9\textwidth]{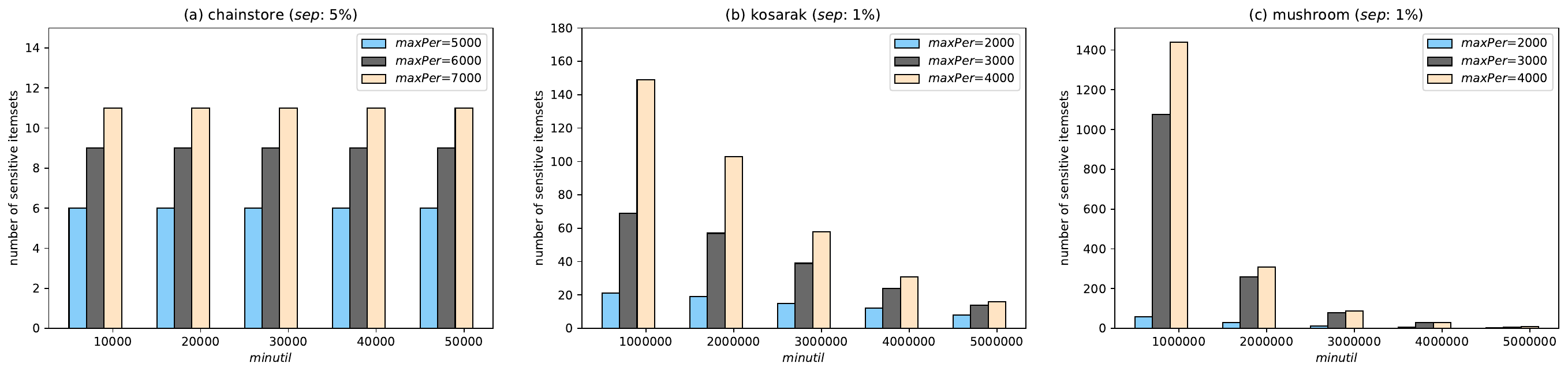} 
  \caption{the number of SPIs under different \textit{minutil} and \textit{maxPer} constraints.}
  \label{fig:numberofspi}
\end{figure*}

As shown in Fig. \ref{fig:numberofspi}, the number of selected sensitive itemsets decreases with the increase of \textit{minutil}. However, when \textit{minutil} reaches a certain threshold, the number of selected itemsets begins to increase with increasing \textit{maxPer}. This observation is consistent with expectations: the number of PHUIs mined from a dataset tends to decrease as \textit{minutil} increases and increases as \textit{maxPer} increases. Since the number of selected sensitive itemsets is a subset of the mined PHUIs, it is naturally affected by these trends. An exception is observed in the chainstore dataset, where the number of selected sensitive itemsets increases with \textit{maxPer} but remains unaffected by changes in \textit{minutil}. This behavior is attributed to the inherent characteristics of the dataset. These findings emphasize the importance of selecting appropriate values for \textit{minutil} and \textit{maxPer}, as the chosen sensitive itemsets directly impact the results of the hiding algorithms.

\begin{figure*}[h!]
  \centering
  \includegraphics[width=0.9\textwidth]{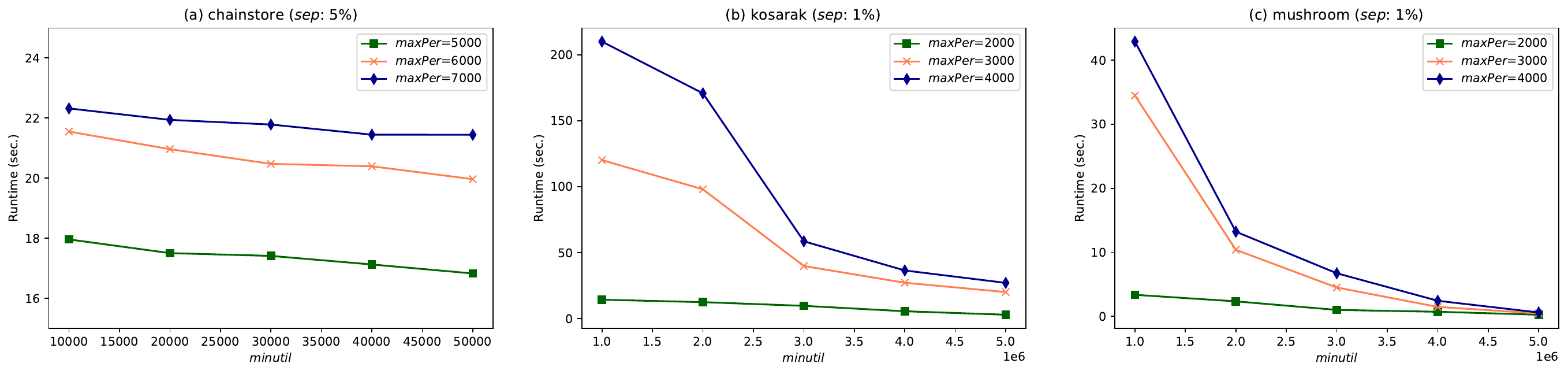} 
  \caption{execution time of MU-MAP with varying \textit{minutil} and \textit{maxPer} settings.}
  \label{fig:runtimesensitivemap}
\end{figure*}

\begin{figure*}[h!]
  \centering
  \includegraphics[width=0.9\textwidth]{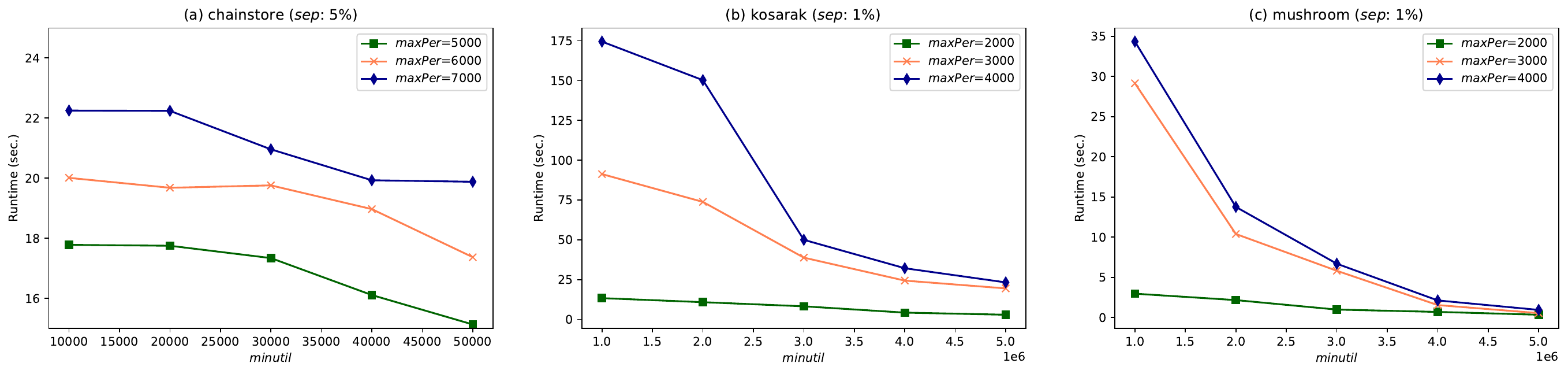} 
  \caption{execution time of MU-MIP with varying \textit{minutil} and \textit{maxPer} settings.}
  \label{fig:runtimesensitivemip}
\end{figure*}

\begin{figure*}[h!]
  \centering
  \includegraphics[width=0.9\textwidth]{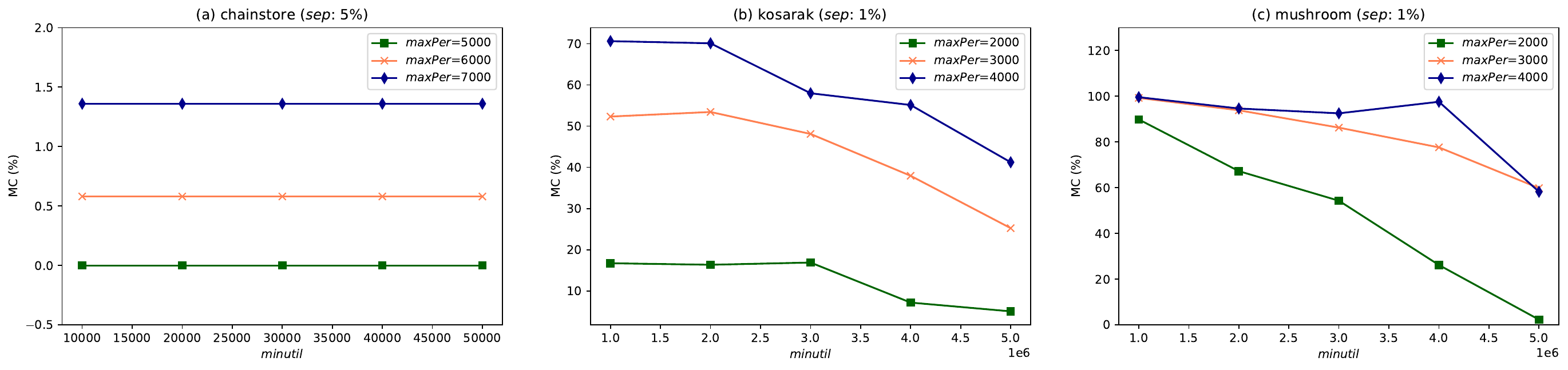} 
  \caption{the MC of MU-MAP with varying \textit{minutil} and \textit{maxPer} settings.}
  \label{fig:mcsensitivemap}
\end{figure*}

\begin{figure*}[h!]
  \centering
  \includegraphics[width=0.9\textwidth]{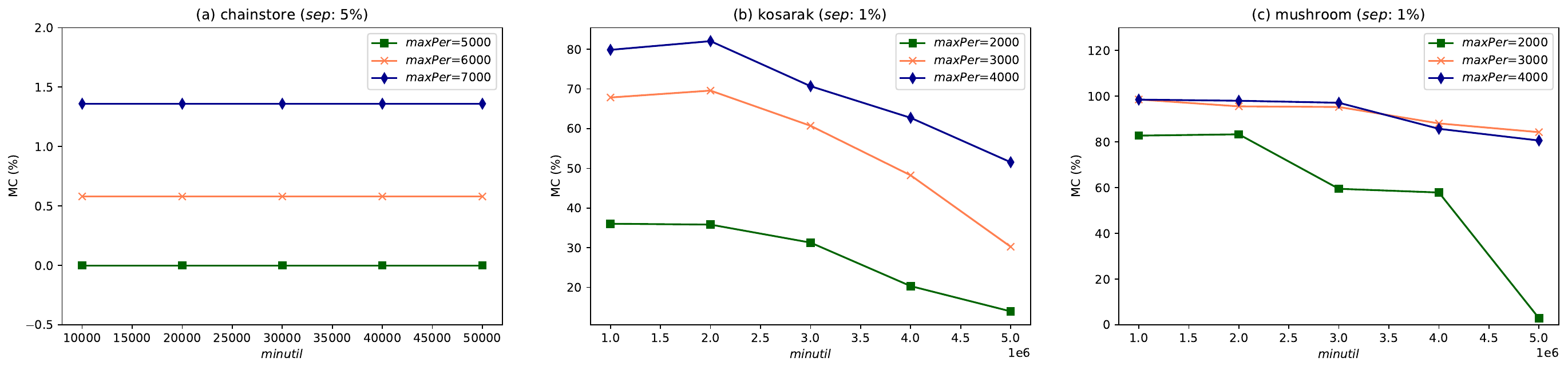} 
  \caption{the MC of MU-MIP with varying \textit{minutil} and \textit{maxPer} settings.}
  \label{fig:mcsensitivemip}
\end{figure*}

To further assess the sensitivity of MU-MAP and MU-MIP to different \textit{minutil} and \textit{maxPer} settings in terms of execution time, additional experiments were conducted. The results are presented in Figures \ref{fig:runtimesensitivemap} and \ref{fig:runtimesensitivemip}. As illustrated, both parameters have a significant impact on algorithm runtime. For example, in Fig. \ref{fig:runtimesensitivemap} (a), using the chainstore dataset, the execution time of MU-MAP with \textit{maxPer} set to 70,000 is approximately 1.05 times that of \textit{maxPer} = 6,000 and about 1.25 times that of \textit{maxPer} = 5,000. A similar trend is observed for MU-MIP in Fig. \ref{fig:runtimesensitivemip} (a). Comparable patterns are also evident in Fig. \ref{fig:runtimesensitivemap} (b)(c), and Fig. \ref{fig:runtimesensitivemip} (b)(c), which correspond to the kosarak and mushroom datasets, respectively. Fig. \ref{fig:mcsensitivemap} and Fig. \ref{fig:mcsensitivemip} show the sensitivity of MU-MAP and MU-MIP to \textit{minutil} and \textit{maxPer} concerning MC. As illustrated, different settings for these parameters significantly influence the MC values. In general, MC decreases as \textit{minutil} increases and increases as \textit{maxPer} increases. This is because a lower \textit{minutil} or a higher \textit{maxPer} typically results in a larger number of sensitive itemsets that must be hidden. However, in the chainstore dataset, MC increases with \textit{maxPer} but remains largely unaffected by changes in \textit{minutil}, since the number of selected sensitive itemsets is relatively insensitive to \textit{minutil} in this dataset. This behavior is a consequence of the dataset's unique characteristics. In summary, varying the values of \textit{minutil} and \textit{maxPer} has a notable impact on the number of selected sensitive itemsets, execution time, and MC. These findings underscore the importance of carefully selecting appropriate values for \textit{minutil} and \textit{maxPer} to ensure the effectiveness and efficiency of the proposed algorithms.

\subsection{Scalability analysis}

\begin{figure}[h!]
  \centering
  \includegraphics[width=0.45\textwidth]{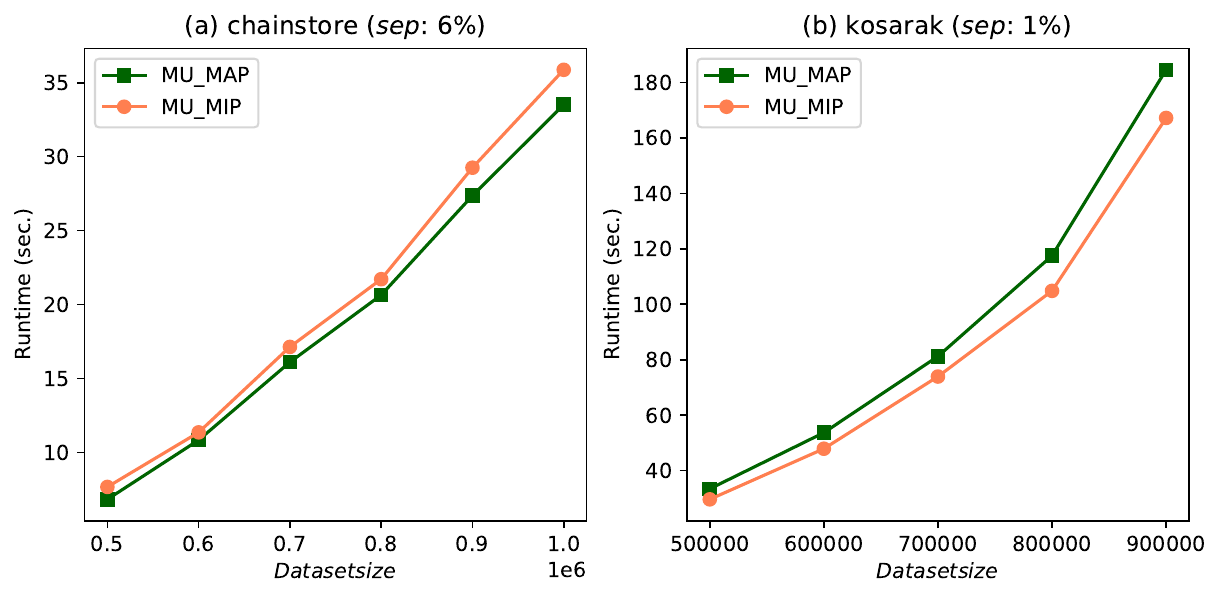} 
  \caption{Runtime under changed dataset sizes.}
  \label{fig:scalability}
\end{figure}

To evaluate the scalability of the MU-MAP and MU-MIP algorithms, we conducted a series of experiments on datasets of varying sizes. The objective was to assess the performance of the algorithms as the number of transactions increases, with a particular focus on comparing execution time, MC, and DSS. We selected two datasets, chainstore and kosarak, for this evaluation. To ensure experimental fairness, the dataset size was the only variable adjusted, while all other parameters were fixed based on the characteristics of each dataset. The periodicity information was adopted from Table \ref{CharacteristicsOfDataset}. Specifically, for kosarak, the \textit{minutil} was set to 5,000,000 and the \textit{sep} to 1$\%$; for chainstore, \textit{minutil} was set to 10,000 and \textit{sep} to 6$\%$. Because the kosarak dataset contains up to 990,002 records, and the chainstore dataset contains up to 1,112,949 records, the size of kosarak varied from 500,000 to 900,000 transactions, while chainstore ranged from 500,000 to 1,000,000 transactions. To further ensure consistency, the selected SPIs for each dataset were kept as similar as possible. The experimental results are illustrated in Fig. \ref{fig:scalability}.

\textbf{Runtime:} As shown in Fig. \ref{fig:scalability}, the execution time of both algorithms increases with the dataset size. This is because a larger dataset results in more transactions being affected by the SPIs, thereby increasing their utility values. Consequently, more transactions must be modified, and a greater amount of utility needs to be reduced to effectively hide the sensitive itemsets, leading to longer execution times. Overall, the growth trend is approximately linear, which demonstrates the good scalability of the proposed algorithms. Moreover, an interesting observation can be made: on the chainstore dataset, the MU-MAP algorithm executes faster than MU-MIP, whereas on the kosarak dataset, MU-MAP takes more time than MU-MIP. This discrepancy can be attributed to the characteristics of the datasets and the selected sensitive itemsets, both of which can significantly impact algorithm performance. If a dataset contains a large number of transactions associated with the sensitive itemsets, the time required for hiding them naturally increases. Additionally, the internal characteristics of the sensitive itemsets can affect performance. For instance, if there is a high degree of item overlap among the sensitive itemsets, modifying one item can simultaneously impact multiple itemsets, potentially accelerating the execution process. \textbf{MC:} As shown in Table \ref{tabMCDatasetSize}, with the increase in dataset size, the MC values of both MU-MAP and MU-MIP remain below 6\% on the chainstore dataset, whereas on kosarak, the MC values are relatively high, exceeding 40\%. This indicates that our algorithms incur a low information loss cost when hiding sensitive itemsets in chainstore, but a relatively higher cost in kosarak. This difference is primarily due to the structural characteristics of the datasets and the distribution of the selected sensitive itemsets within them. \textbf{DSS:} As shown in Table \ref{tabdssdatasetsize}, as the dataset size increases, the DSS values for both MU-MAP and MU-MIP remain above 98\% on both chainstore and kosarak. This demonstrates that the proposed algorithms are able to preserve a high degree of structural similarity in the datasets, even after hiding the sensitive itemsets. In summary, the scalability analysis demonstrates that both MU-MAP and MU-MIP exhibit strong scalability as the dataset size increases.

\begin{table}[htbp]
    \footnotesize
    \centering
    \caption{The MC (\%) under different dataset sizes.}
\resizebox{8cm}{!}{ 
\begin{tabular}{|c|ccccccc|}
\hline
\textbf{Dataset} & \textbf{Parameter} & \textbf{test$_1$} & \textbf{test$_2$} & \textbf{test$_3$} & \textbf{test$_4$} & \textbf{test$_5$} & \textbf{test$_6$}\\ \hline

\multirow{1}{*}{chainstore} & \textit{dataset size($\times$ $10^4$)} &5& 6 & 7 & 8 & 9 & 10  \\ \cline{2-8}
&MU-MAP&4.90& 4.72 & 5.39 & 5.29 & 5.49 & 4.55 \\  
&MU-MIP&4.49& 4.29 & 4.90 & 4.76 & 4.95 & 4.55\\  \hline

\multirow{1}{*}{kosarak} & \textit{dataset size($\times$ $10^4$)} &5& 6 & 7 & 8 & 9 & /  \\ \cline{2-8}
&MU-MAP&40.15& 46.47 & 46.41 & 46.41 & 48.87 & /  \\  
&MU-MIP&41.61& 44.77 & 45.24 & 47.63 & 52.52 & / \\  \hline

\end{tabular}
}
\label{tabMCDatasetSize}
\end{table}

\begin{table}[htbp]
    \footnotesize
    \centering
    \caption{The DSS (\%) under different dataset sizes.}
\resizebox{8cm}{!}{ 
\begin{tabular}{|c|ccccccc|}
\hline
\textbf{Dataset} & \textbf{Parameter} & \textbf{test$_1$} & \textbf{test$_2$} & \textbf{test$_3$} & \textbf{test$_4$} & \textbf{test$_5$} & \textbf{test$_6$}\\ \hline

\multirow{1}{*}{chainstore} & \textit{dataset size($\times$ $10^4$)} &5& 6 & 7 & 8 & 9 & 10  \\ \cline{2-8}
&MU-MAP&99.11& 99.10 & 99.18 & 99.23 & 98.23 & 99.28 \\  
&MU-MIP&99.11& 99.10 & 99.18 & 99.23 & 98.23 & 99.28\\  \hline

\multirow{1}{*}{kosarak} & \textit{dataset size($\times$ $10^4$)} &5& 6 & 7 & 8 & 9 & /  \\ \cline{2-8}
&MU-MAP&99.88& 99.86 & 99.84 & 99.83 & 99.82 & /  \\  
&MU-MIP&99.88& 99.86 & 99.85 & 99.84 & 99.83 & / \\  \hline

\end{tabular}
}
\label{tabdssdatasetsize}
\end{table}

Through the discussion and analysis in the experimental section, we found that both the distribution of items in the dataset and the density of the dataset affect the periodicity of itemsets, which in turn impacts the algorithm's performance. The MU-MAP algorithm is more suitable for sparse datasets, where the item distribution is more dispersed and the utility values of items are relatively small. On the other hand, the MU-MIP algorithm is more suitable for dense datasets, where the item distribution is more concentrated and the utility values of items are higher.

\section{Conclusion}  \label{sec: conclusion}

With the advent of the big data era, privacy preservation has become an essential and pressing research topic. PPUM addresses privacy leakage during HUIM. However, existing PPUM approaches are not suitable for domains that require hiding periodic sensitive information, such as smart home systems and healthcare environments. To bridge this gap, we introduced a novel research task, PPPUM, and proposed two effective algorithms, MU-MAP and MU-MIP, specifically designed to hide sensitive PHUIs while maintaining data utility. To improve mining efficiency and performance, we developed two novel data structures, SISL and SIL. We conducted extensive experiments on six real-world datasets, comparing our algorithms with three state-of-the-art PPUM methods (SMAU, SMIU, and SMSE). Evaluation was based on five widely used metrics: IUS, DUS, DSS, AC, and MC. Experimental results demonstrate that MU-MAP and MU-MIP can effectively hide sensitive periodic itemsets while avoiding the generation of AC. In contrast, SMAU, SMIU, and SMSE introduce AC when addressing the PPPUM problem. Furthermore, we analyzed the impact of sensitive itemset selection strategies and found that our algorithms outperform traditional PPUM approaches under incremental sensitive itemset scenarios. We also conducted significance analysis, sensitivity analysis, and scalability experiments to rigorously validate the robustness and effectiveness of our proposed methods.

Although the proposed algorithms achieve a promising balance between privacy preservation and data utility, there remains room for improvement. Future work will explore more refined hiding strategies and optimized data structures to further enhance algorithmic efficiency and scalability. In particular, future research will focus on evaluating the applicability of the proposed methods under two critical conditions: (1) ultra-long periodicity scenarios, where patterns span extensive time intervals, and (2) large-scale datasets, which pose significant computational and memory challenges. Moreover, we plan to extend our approach to support dynamic data stream environments and distributed computing frameworks. These extensions will require addressing key challenges such as real-time processing requirements, incremental model updates, and computational complexity. Through these efforts, we aim to further enhance the practicality and robustness of privacy-preserving utility mining in real-world applications.

\bibliographystyle{IEEEtran}
\bibliography{pppum}

\end{document}